\documentclass[twocolumn]{aastex631}

\usepackage{amsmath}

\begin{document}

\title{Exploring X-ray Properties of Low Metallicity Dwarf Galaxies}

\author[0000-0003-1051-6564]{Jenna M. Cann}
\affiliation{X-ray Astrophysics Laboratory, NASA Goddard Space Flight Center, Code 662, Greenbelt, MD 20771, USA}
\affiliation{Oak Ridge Associated Universities, NASA NPP Program, Oak Ridge, TN 37831, USA}

\author{Kimberly A. Weaver}
\affiliation{X-ray Astrophysics Laboratory, NASA Goddard Space Flight Center, Code 662, Greenbelt, MD 20771, USA}

\author[0000-0001-8640-8522]{Ryan W. Pfeifle}
\affiliation{X-ray Astrophysics Laboratory, NASA Goddard Space Flight Center, Code 662, Greenbelt, MD 20771, USA}
\affiliation{Oak Ridge Associated Universities, NASA NPP Program, Oak Ridge, TN 37831, USA}

\author[0000-0002-4902-8077]{Nathan J. Secrest}
\affiliation{U.S. Naval Observatory, 3450 Massachusetts Avenue NW, Washington, DC 20392, USA}

\author{Shobita Satyapal}
\affiliation{Department of Physics and Astronomy, George Mason University, 4400 University Drive, MSN 3F3, Fairfax, VA 22030, USA}

\author{Mario Gliozzi}
\affiliation{Department of Physics and Astronomy, George Mason University, 4400 University Drive, MSN 3F3, Fairfax, VA 22030, USA}

\begin{abstract}

One of the primary outstanding questions in extragalactic astronomy is the formation and early evolution of the supermassive black holes that are seen in nearly every massive galaxy. 
Low metallicity dwarf galaxies may offer the most representative local analogs to pristine early galaxies, making them a vital tool in probing black hole seed models through the study of the intermediate mass black holes (IMBHs) possibly hosted therein, though these dwarf galaxies, and the IMBHs they may host, are typically not as well-studied in this context as their higher metallicity and higher mass counterparts. In this paper, we explore the X-ray properties of a sample of 37 low metallicity dwarf galaxies using archival XMM observations, and we compare the properties of this population against a representative sample of higher metallicity counterparts. We report the detection of ten sources with 0.3-10 keV luminosity in excess of $10^{40}$~erg~s$^{-1}$ within the low metallicity sample, which we highlight for follow-up as potential intermediate mass black hole candidates. Finally, we discuss the differing multi-wavelength scaling relations (e.g., $L_X - L_{W2}$, $L_X-SFR$) between the two galaxy populations, as well as the sample's $L_X$ as a function of metallicity.



\end{abstract}

\keywords{Dwarf galaxies, X-ray Astronomy, Active Galaxies}

\section{Introduction} \label{sec:intro}

In recent years, dedicated study of active galactic nuclei (AGNs) in dwarf galaxies has greatly increased our knowledge of this population, and expanded the range of black hole masses known to reside in the centers of galaxies down to $\approx 10^4$\,M$_\odot$ \citep[e.g., see review by][]{greene2020}. One significant motivation for this work is the search for intermediate mass black holes (IMBHs; $\approx100 - 10^4$\,M$_\odot$). These objects have garnered significant scientific interest due to their potential to provide insight into the origins of supermassive black holes: their mass function and galaxy occupation fraction are theorized to provide constraints into the formation and evolution models of seed black holes formed at high redshift \citep[e.g.,][]{volonteri2010,reines2016}. 

The premise behind this potential is that the IMBHs in dwarf galaxies have had a relatively quiescent cosmic history, as mergers and interactions with other galaxies may cause the central black hole to undergo a different evolution than it would through purely in situ processes. For example, mergers and interactions have been shown to drive gas to the center of the galaxy, kickstarting star formation, enriching the gas metallicity, and potentially growing the central supermassive black hole beyond that expected through a more quiescent cosmic history \citep{barnes1996,dimatteo2005,hopkins2008}. The information provided by observations of these black holes may therefore bias the derived seed models towards more massive seeds in the absence of information from a potentially unexplored population \citep{reines2016}. For this reason, there has been renewed interest in ``pristine" galaxies, that have evolved relatively quietly from their initial formation, and can be considered our closest local analogs to the high redshift galaxies that hosted the first black hole seeds. To probe this population necessitates looking not only at low mass galaxies, but specifically also at low metallicity dwarf galaxies \citep[see][for a review]{mezcua2019}, which have been less well-studied in this aspect than their higher metallicity counterparts.  

The majority of studies dedicated to low metallicity dwarf galaxies have focused on sources that are presumed to not host an AGN. Several multi-wavelength studies have been done on the lowest metallicity ($Z<\approx10\%$ Solar) blue compact dwarfs (BCDs) in large surveys \citep[e.g.,][]{izotov2012b,izotov2015,izotov2016,thuan2016,guseva2017,izotov2019a,guseva2019} or in-depth studies of individual objects of interest \citep[e.g.,][]{cairos2010,izotov2018,izotov2019b,izotov2021a,izotov2021b,thuan2022,ju2022}, including studies of eleven BCDs that display [\ion{Ne}{5}] coronal line emission \citep{izotov2012,izotov2021b}. In the X-rays, a theoretical study of X-ray binary (XRB) populations in low metallicity galaxies suggests that the X-ray emission from XRBs can be enhanced by over an order of magnitude when compared to higher metallicity galaxies \citep{kaaret2011,fragos2013,brorby2014} and \citet{prestwich2013} shows that ultra-luminous X-ray sources can be more common in low metallicity galaxies, both of which could have X-ray luminosities comparable to, and indistinguishable from, IMBH-powered AGNs. A study of ULXs hosted in low metallicity galaxies suggests a potential anti-correlation between metallicity, star-forming rate, and number of ULXs or black holes \citep{mapelli2010}. Chandra studies by \citet{lehmer2021,lehmer2022} of low metallicity galaxies with stellar masses $\approx10^8-10^{9.5}$~M$_\odot$ and $\approx10^{8.5}-10^{10.5}$~M$\odot$, respectively, explored the  star formation rate (SFR)-$L_X$ relation of this population, and they found agreement with previous theoretical results \citep{fragos2013,lehmer2019} suggesting systematically higher X-ray luminosities from stellar sources in low metallicity galaxies when compared to their higher metallicity counterparts. These studies, while not focused on AGN activity, hold important results for AGN searches in this population, as they unveil significant emission from stellar sources that may be incorrectly attributed to AGN activity.

%


There have also been some low metallicity studies that do focus on AGN activity. For example, \citet{groves2006} analyzed $\approx23,000$ galaxies in the Sloan Digital Sky Survey (SDSS) to determine emission line diagnostics indicative of low metallicity AGNs, and found only 40 candidates. \citet{izotov2007,izotov2008,izotov2010,burke2021} studied the optical spectra of low metallicity galaxies and found broad line emission indicative of AGNs. In a multi-wavelength study, \citet{cann2020} explored X-ray, infrared, and optical properties of a low metallicity ($\approx10\%$ Solar) AGN, and uncovered that the AGN had an X-ray luminosity several orders of magnitude below that expected from its mid-infrared luminosity based on the $L_{W2}-L_{2-10 keV}$ relation in \citet{secrest2015}. This result can also be seen in \citet{dong2012} and \citet{simmonds2016}, in which AGN candidates identified through broad H$\alpha$ emission in low metallicity hosts show a similar systematic X-ray deficit compared to that expected from their $l2500$ \citep{steffen2006} and [\ion{O}{3}] \citep{panessa2006} luminosities, respectively. There have also been several recent IMBH searches that include or target low metallicity dwarf galaxies \citep[e.g.,][]{simmonds2016,cann2020,cann2021,molina2021,reefe2022a} using a variety of AGN identification techniques. The use of optical \citep[e.g.,][]{izotov2012,molina2021, reefe2022a} and near-infrared \citep{cann2021} coronal lines have unearthed AGN candidates in low metallicity galaxies with no other evidence for AGN emission. \citet{thygesen2023} uses Chandra, VLA, and HST data to uncover a potential AGN candidate in low metallicity galaxy, Mrk 1434.

With the increased scientific interest on searching for AGNs in low metallicity dwarf galaxies, further study is needed on the emission from this population. As noted above, there can be considerable ambiguity regarding the source of high energy emission in these galaxies, which results in large uncertainties in the identification of potential candidate AGNs, hampering progress in answering the fundamental questions listed above regarding the origins of supermassive black holes. In this work, we present a thorough investigation of the X-ray properties of low metallicity dwarf galaxies. In Section~\ref{sec:methods}, we describe our sample selection and data reduction. In Section~\ref{sec:results}, we present our results, including the tabulated count rates and luminosities of the entire low metallicity sample and an exploration of the X-ray spectra of one case-study galaxy. In Section~\ref{sec:implications}, we present luminosity relations of this sample compared to current literature, discuss in more detail specific high luminosity sources in our sample, and highlight implications for current and future missions. We adopt a standard $\Lambda$CDM cosmology with $H_0=70$~km~s$^{-1}$~Mpc$^{-1}$, $\Omega_{\text{M}}=0.3$, and $\Omega_\Lambda=0.7$.

\section{Methodology} \label{sec:methods}

\subsection{Sample Selection}
An initial sample of low metallicity dwarf galaxies observed by the Sloan Digital Sky Survey (SDSS) was compiled using the Max Planck Institut f\"{u}r Astrophysik/Johns Hopkins
University (MPA/JHU) catalog for SDSS~DR8. To select local dwarf galaxies, we required a mass upper limit of $\log M_* < 9.5$, which is approximately the mass of the Large Magellanic Cloud (LMC), and a redshift cut of $z < 0.1$.  We also required a metallicity cut of $\log$[\ion{N}{2}]/H$\alpha < -1.0$, which corresponds to sub-Solar metallicities \citep{groves2006}, and required the signal-to-noise ratio of the [\ion{N}{2}] and H$\alpha$ lines to be greater than 3 to ensure robust measurements.  
This results in an initial sample of $\approx$11,000 galaxies. This sample was then crossmatched with the XMM-Newton Observation Log within 15$\arcmin$ to find XMM observations that may have serendipitously observed objects in our sample.  This resulted in a sample of 99 objects, which were then visually inspected to ensure that they were in fact dwarf galaxies, and not \ion{H}{2} regions of larger galaxies.  After this procedure the final sample consisted of 37 galaxies with metallicities ranging from $\approx 15-40\%$ Solar (Table~\ref{sample_demo}). 

\begin{table*}[t]
    \caption{The low metallicity dwarf galaxy sample}
    \centering
    \begin{tabular}{cccccccccc}
    \hline
    Target & RA & Dec & $z$ & $\log(M_*/M_\odot)$ & $\log$([\ion{N}{2}]/H$\alpha$) & $12+\log$(O/H) & Exp. Time \\
    &  &  & &  & &  & ksec\\
    \noalign{\smallskip}
    \hline
    \noalign{\smallskip}
    J014529.37+001039.4 & 26.3724 & 0.1776 & 0.0805 & 9.06 & -1.28 & 8.17 & 14.0\\
    J015314.18+010616.7 & 28.3091 & 1.1047 & 0.0562 & 9.21 & -1.08 & 8.25 & 5.1 \\
    J023026.91+004839.7 & 37.6122 & 0.8111 & 0.0413 & 8.96 & -1.08 & 8.24 & 6.7 \\
    J025645.59+060318.4 & 44.1900 & 6.0551 & 0.0262 & 6.49 & -1.28 & 8.17 & 6.7 \\
    J030613.83-002037.9 & 46.5581 & -0.3436 & 0.0295 & 9.19 & -1.16 & 8.21 & 2.7 \\
    J082228.93+034551.7 & 125.6206 & 3.7643 & 0.0351 & 8.82 & -1.11 & 8.23 & 35.0 \\
    J082527.65+295739.3 & 126.3652 & 29.9609 & 0.0503 & 9.02 & -1.13 & 8.22 & 23.9 \\
    J082556.54+040651.6 & 126.4856 & 4.1143 & 0.0506 & 8.78 & -1.01 & 8.27 & 21.1 \\
    J083200.50+191205.8 & 128.0021 & 19.2016 & 0.0375 & 9.29 & -1.57 & 8.05 & 19.9 \\
    J084220.94+115000.3 & 130.5873 & 11.8334 & 0.0294 & 8.33 & -1.42 & 8.11 & 12.0 \\
    J085834.17+135541.7 & 134.6424 & 13.9283 & 0.0169 & 8.54 & -1.03 & 8.27 & 116.5 \\
    J100710.97+123905.0 & 151.7957 & 12.6514 & 0.0093 & 8.23 & -1.17 & 8.21 & 24.7 \\
    J102400.56+195927.5 & 156.0024 & 19.9910 & 0.0473 & 8.91 & -1.08 & 8.24 & 59.6 \\
    J103410.15+580349.1 & 158.5423 & 58.0636 & 0.0070 & 7.00 & -1.86 & 7.89 & 5.0 \\
    J112852.09+171604.4 & 172.2171 & 17.2679 & 0.0112 & 8.02 & -1.09 & 8.24 & 32.5 \\
    J113908.83+193459.8 & 174.7868 & 19.5833 & 0.0058 & 7.12 & -1.37 & 8.13 & 32.0 \\
    J115814.05+261834.0 & 179.5585 & 26.3095 & 0.0231 & 8.57 & -1.24 & 8.18 & 45.3 \\
    J120013.14+315156.1 & 180.0548 & 31.8656 & 0.0104 & 7.88 & -1.15 & 8.22 & 7.9 \\
    J121345.22+363805.8 & 183.4365 & 36.6340 & 0.0007 & 7.06 & -1.33 & 8.14 & 37.8 \\
    J122106.01+113543.2 & 185.2753 & 11.5953 & 0.0031 & 7.71 & -1.18 & 8.20 & 117.0 \\
    J122140.31+113011.3 & 185.4180 & 11.5031 & 0.0008 & 6.22 & -1.03 & 8.27 & 16.4 \\
    J122809.20+440508.16 & 187.0386 & 44.0856 & 0.0007 & 6.27 & -1.08 & 8.24 & 41.0 \\
    J123520.04+393109.0 & 188.8335 & 39.5192 & 0.0209 & 7.90 & -1.77 & 7.95 & 16.6 \\
    J123551.80+275557.00 & 188.9661 & 27.9325 & 0.0027 & 6.18 & -1.62 & 8.03 & 109.4 \\
    J124141.20+332447.0 & 190.4217 & 33.4130 & 0.0228 & 8.75 & -1.21 & 8.19 & 116.2 \\
    J125900.27+345044.8 & 194.7517 & 34.8465 & 0.0027 & 6.79 & -1.66 & 8.01 & 9.1 \\
    J130728.44+542652.5 & 196.8685 & 54.4479 & 0.0326 & 8.79 & -1.19 & 8.20 & 45.8 \\
    J130821.42+113055.0 & 197.0893 & 11.5153 & 0.0250 & 8.31 & -1.32 & 8.15 & 12.5 \\
    J132030.21+330842.8 & 200.1259 & 33.1452 & 0.0182 & 8.77 & -1.08 & 8.24 & 55.0 \\
    J132844.05+435550.5 & 202.1836 & 43.9307 & 0.0280 & 8.34 & -1.25 & 8.18 & 27.1 \\
    J132932.41+323416.9 & 202.3851 & 32.5714 & 0.0156 & 8.08 & -1.05 & 8.26 & 63.7 \\
    J134427.36+560129.7 & 206.1140 & 56.0249 & 0.0706 & 9.15 & -1.13 & 8.22 & 25.6 \\
    J142054.72+395451.1 & 215.2280 & 39.9142 & 0.0056 & 7.14 & -1.19 & 8.20 & 40.0 \\
    J142459.39+225448.0 & 216.2475 & 22.9134 & 0.0331 & 8.62 & -1.06 & 8.25 & 118.9 \\
    J154848.17+351718.6 & 237.2008 & 35.2885 & 0.0552 & 8.64 & -1.21 & 8.19 & 26.0 \\
    J160651.01+080928.6 & 241.7126 & 8.1580 & 0.0093 & 8.21 & -1.01 & 8.27 & 1.5 \\
    J212855.85+000325.6 & 322.2328 & 0.0571 & 0.0309 & 7.91 & -1.37 & 8.13 & 42.8 \\
    \noalign{\smallskip}
    \hline
    \noalign{\smallskip}
    \end{tabular}
     {Coordinates, redshifts, masses \citep{kauffmann2003}, and $\log$([N II]/H$\alpha$) values of the low metallicity dwarf sample from dr8 of the MPA/JHU catalog. The exposure time column reports the good time interval of the images. Metallicity estimates were calculated using \citet{pettini2004}.}
    \label{sample_demo}
\end{table*}

A control sample to compare gross properties with the target sample was also created following the same procedure and the same constraints in mass, redshift, and SNR, but requiring $\log$[\ion{N}{2}]/H$\alpha > -1.0$, resulting in 207 galaxies. Figure 1 displays a comparison of the physical properties of the target and control samples.  The star-forming rates and masses of the control and sample populations were compared using the Anderson-Darling Test, which tests the null hypothesis that the two samples arise from the same parent population.  The null hypothesis could not be rejected at the 95\% confidence level for all quantity comparisons, thus we cannot rule out that they arise from the same parent population.  We can therefore statistically compare the two samples. \textbf{We emphasize that there was no intentional selection for AGN or star-formation dominated galaxies, however we
note that the choice of metallicity diagnostic results in a slightly higher proportion of BPT-identified AGNs in the control sample. For a more complete discussion of the potential caveats and biases resulting from this sample selection, see section 4.4.}

\begin{figure}[h]
\centering
\includegraphics[width=0.99\linewidth]{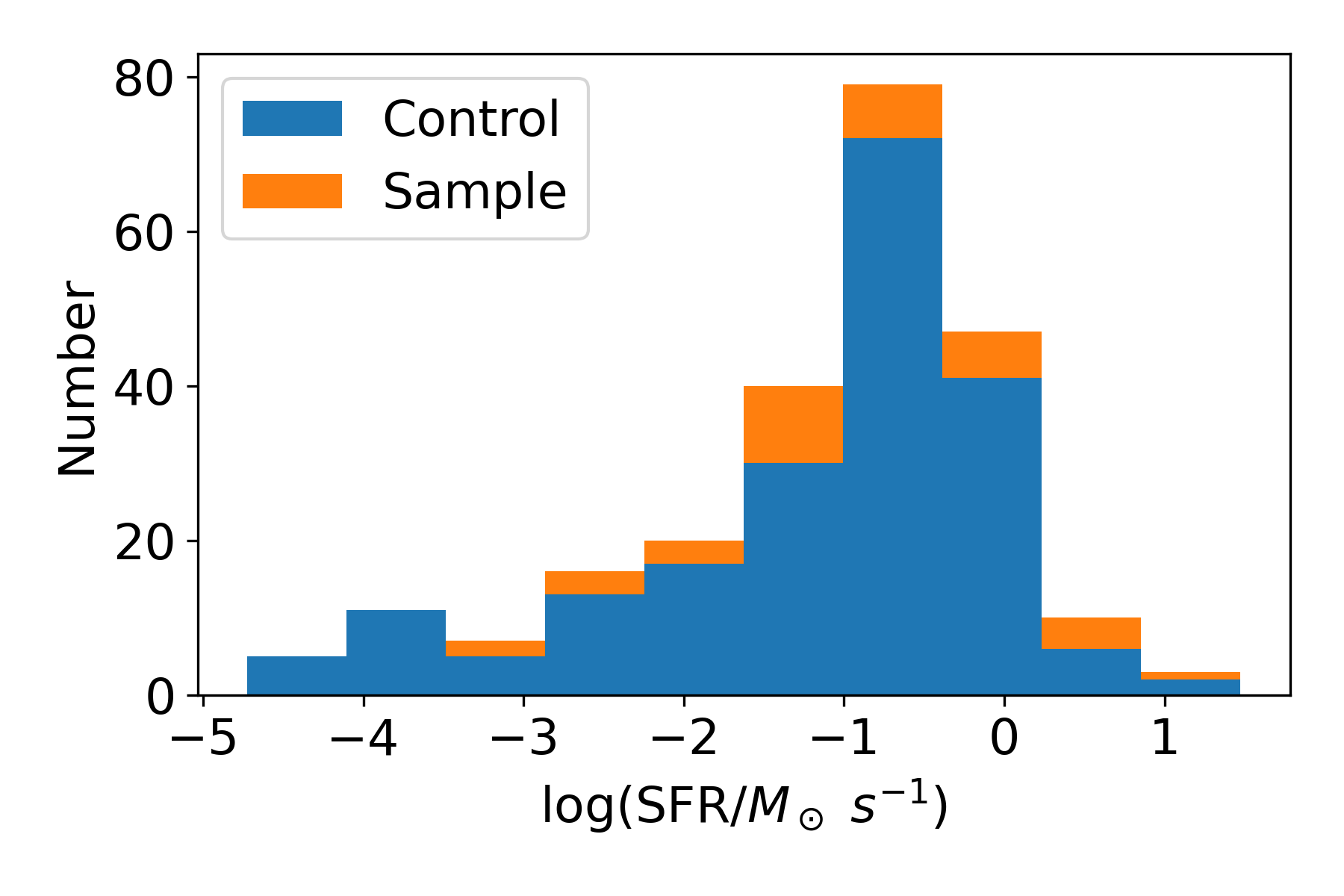} \\
\includegraphics[width=0.99\linewidth]{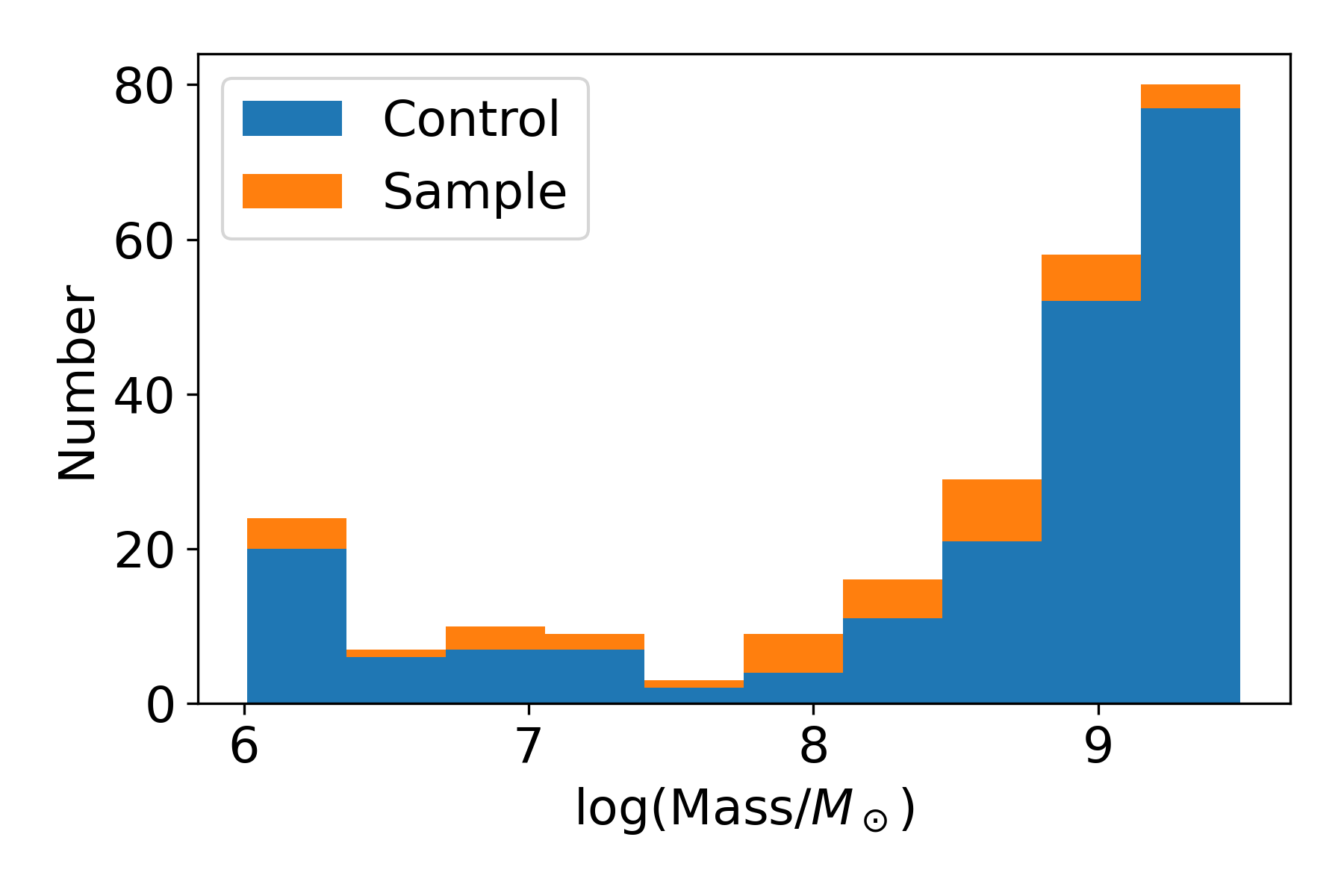}\\
\includegraphics[width=0.99\linewidth]{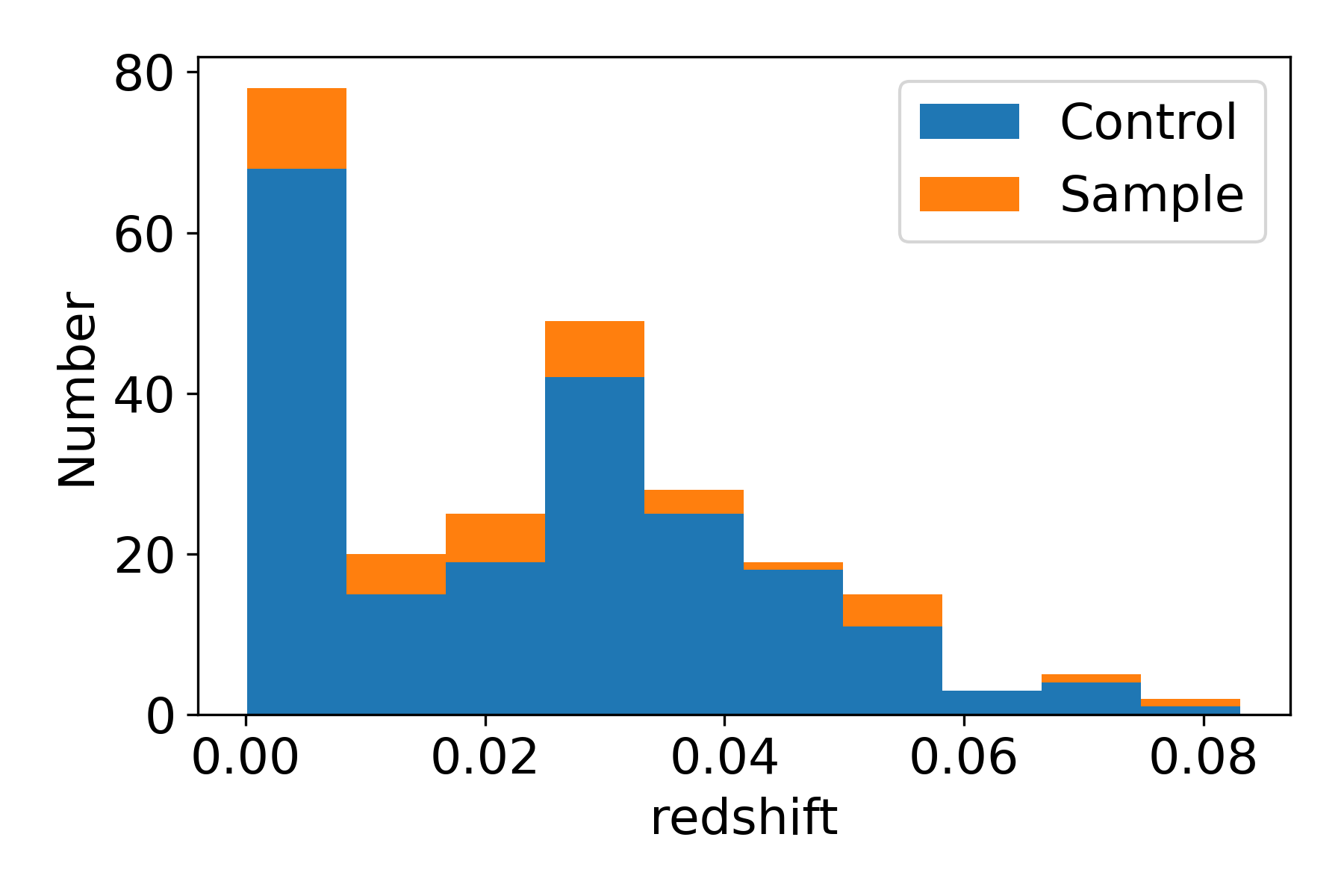}\\ 
\caption{\footnotesize
The distribution of star formation rates \citep[top; ][]{brinchmann2004, salim2007}, stellar masses \citep[middle; ][]{kauffmann2003}, and redshifts ($z$) of the low metallicity dwarf sample compared with the control sample, as reported by the MPA-JHU catalog.}
\label{fig1}
\end{figure}



\subsection{X-ray Data Reduction}

The X-ray data were reduced using version 19.1.0 of the XMM Source Analysis Software (SAS). A custom Python code, adapted from \citet{secrest2020}, was used to download, reprocess and filter the data, and extract source counts and 3$\sigma$ upper limits. This standardized approach allows us to directly compare the archival observations and uniformly judge source detections across multiple observation cycles. 

For each ObsID, the pipeline calls cifbuild and odfingest to produce a CCF and SAS summary file, respectively, for the observation.  The pn and MOS event files are then produced using epchain and emchain. When possible to optimize the sensitivity of the sample, we took advantage of the presence of both scheduled and unscheduled exposures and combined them into single event files using evlistcomb. These event files were then filtered for science-grade events using evselect and requiring PATTERN$<$4 for the pn and PATTERN$<$12 for the MOS. High energy background light curves with (PI$>$10000\&\&PI$<$12000) for pn and (PI$>10000$\&\&PI$<15000$) for MOS were created, and periods of high background when the total count rate exceed 0.35 cps for pn and 0.40 cps for MOS were removed. Hard (2--10 keV), soft (0.3--2 keV), and full (0.3--10 keV) band images were produced using evselect, and source counts were extracted using a $15\arcsec$ radius aperture positioned at the SDSS coordinates for the optical center of the galaxy from each image using eregionanalyse. Background counts were extracted from a source-free region in the FOV using a $60\arcsec$ radius aperture, with off-axis vignetting accounted for by supplying eregionanalyse with an exposure map generated using eexpmap. The exposure map is also used to ensure that the source region does not fall within a chip gap or other area with censored coverage. Rest-frame fluxes were obtained from count rates using a power law model with $\Gamma = 1.7$ \citep[see][for more information]{secrest2020}, with revised energy conversion factor (ECF) look-up tables for each filter using version 4.12a of the Portable, Interactive Multi-Mission Simulator \citep[PIMMs;][]{mukai1993} via WebPIMMs. 

The detector background in XMM varies spatially, and so with an automated pipeline, it is possible in a small number of sources that the source can be over-subtracted, due to the automatically defined background regions. After the pipeline was run, we examined all upper limits for signs of over-subtraction (i.e., those that had negative flux values). For these sources, we could not trust that the upper limit provided by the pipeline was accurate and we recalculated the upper limits manually. To calculate accurate upper limits, we derived a new background and background-subtracted spectrum in \textsc{Xspec}, and then determined the normalization of an absorbed power law model that best fit that spectrum. We then calculated the flux corresponding to that normalization, thereby determining the upper limit by which a power law component with $\Gamma=1.7$ would be detected in the data. For sources identified as 3$\sigma$ detections, source and background spectra were extracted using the evselect command, with corresponding response files generated using the arfgen and rmfgen commands.

\section{Results} \label{sec:results}

\subsection{X-ray Fluxes}

\begin{figure}[h]
\centering
\includegraphics[width=0.99\linewidth]{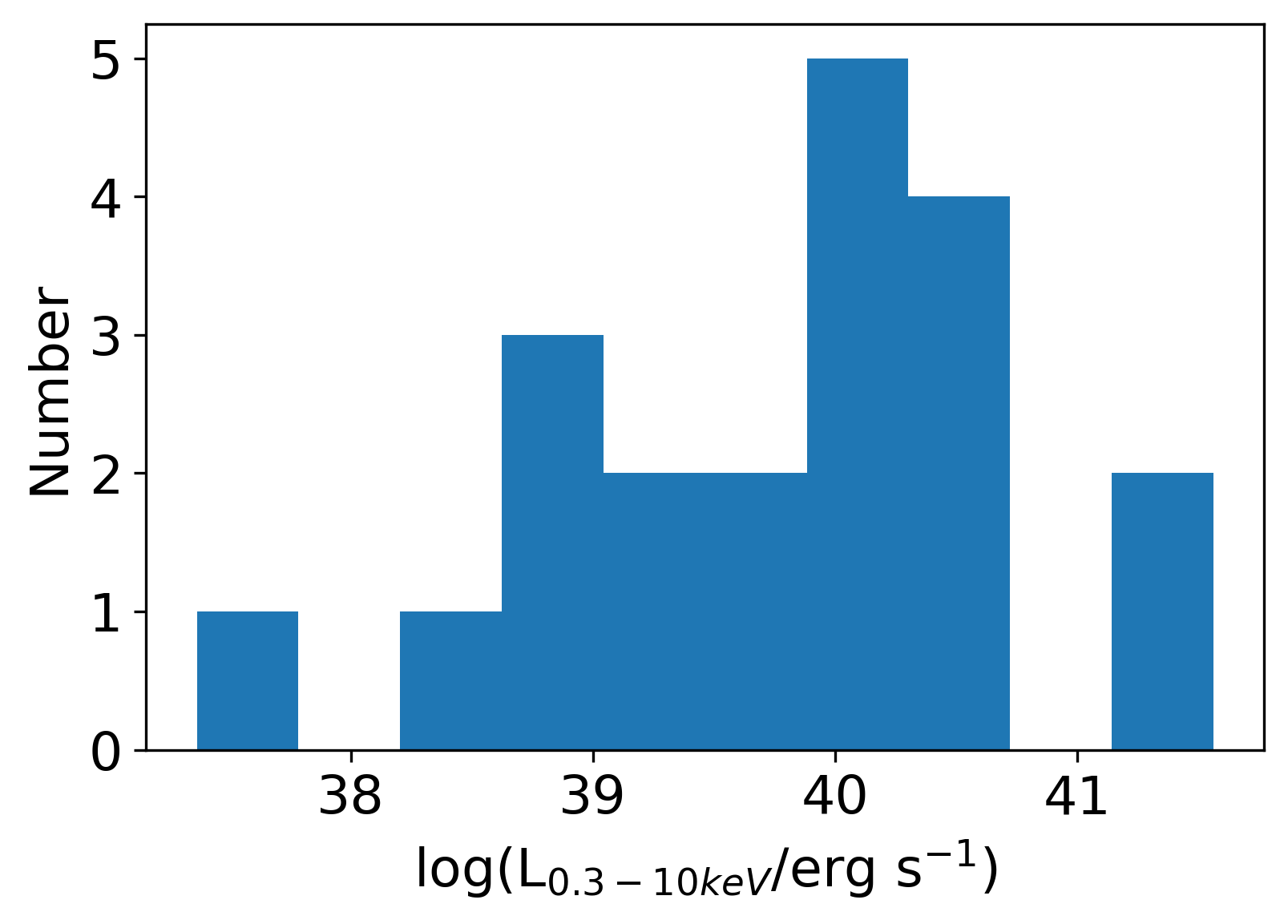} \\
\caption{\footnotesize A histogram showing the 0.3--10\,keV X-ray luminosities for the detected ($>3\sigma$) sources in the low metallicity dwarf sample.} 
\label{fig2}
\end{figure}

The X-ray fluxes and 3$\sigma$ upper limits of the sample are listed in Table 2. We note that for this archival study, that includes both serendipitous and pointed observations, we had a 52.6\%, 26.3\%, and 52.6\% detection rate in the full, hard, and soft bands, respectively. By comparison, the control sample had a 49.7\%, 26.1\%, and 49.2\% detection rate in the full, hard, and soft bands, very similar to the low metallicity sample. A histogram of the full band luminosities of the low metallicity sample can be found in Figure~\ref{fig2}. Note the presence of several sources with X-ray luminosities above $\approx10^{40}$erg~s$^{-1}$, which is typically indicative of an AGN in the low mass regime \citep{greene2007,cann2018}. A detailed analysis of these sources can be found in Section \ref{sec:implications}. For sources with well-defined ($3\sigma$) detections in both the hard and soft bands, we also quote a hardness ratio, defined as (H-S)/(H+S), where H and S are the count rates in the hard (2-10 keV) and soft (0.3-2 keV) bands respectively, though there are insufficient detections to analyze trends compared to the control sample in detail. In short, the control sample spanned a much wider range of hardness ratios, and had a significantly higher median hardness ratio compared to the low metallicity sample (0.22 vs. 0.085).

\begin{table*}[t]
    \centering
    \caption{Count rates, fluxes, and luminosities of the low metallicity dwarf sample}
    \begin{tabular}{ccccccc}
    \hline
    \noalign{\smallskip}
    Target & Ct rate (0.3--10\,keV) & $F_\textrm{0.3--10\,keV}$ & $F_\textrm{2--10\,\rm{keV}}$ &  L$_\textrm{0.3--10\,\rm{keV}}$ &  L$_\textrm{2--10\,\rm{keV}}$ &  HR$^*$ \\
& $10^{-3}$cps  & $10^{-14}$\,erg~cm$^{-2}$~s$^{-1}$ & $10^{-14}$\,erg~cm$^{-2}$~s$^{-1}$ & $\log{L/}$\,erg~s$^{-1}$ & $\log{L}$\,erg~s$^{-1}$ & \\
    \noalign{\smallskip}
    \hline
    \noalign{\smallskip}
    J0145+0010 & $8.32\pm1.83$ & $2.29\pm0.50$ & $<1.67$ & $41.56 \pm 0.43$ & $<41.43$ & \dots \\
    J0153+0106 & $<5.49$ & $<1.51$  &$<3.18$ & $<41.06$ & $<41.38$ & \dots \\
    J0230+0048 & $<6.15$ & $<1.68$  & $<3.68$ & $<40.83$ & $<41.17$ & \dots \\
    J0256+0603 & $<6.30$ & $<1.76$  & $<3.65$ & $<40.44$ & $<40.76$ & \dots \\
    J0306-0020 & $<9.45$ & $<2.58$  & $<4.78$ & $<40.71$ & $<40.98$ & \dots \\
    J0822+0345 & $<3.45$ & $<0.95$ & $<1.53$ & $<40.43$ & $<40.64$ & \dots \\
    J0825+2957 & $<3.44$ & $<0.94$ & $<1.44$ & $<40.75$ & $<40.93$ & \dots \\
    J0825+0406 & $<3.99$ & $<1.09$ & $<1.74$ & $<40.82$ & $<41.02$ & \dots \\
    J0832+1912 & $4.83\pm1.27$ & $1.33 \pm 0.35$ & $<1.53$ & $40.64 \pm 0.43$ & $<40.70$ & \dots \\
    J0842+1150 & $<4.18$ & $<1.14$ & $<2.56$ & $<40.36$ & $<40.71$ & \dots \\
    J0858+1355 & $2.25\pm0.51$ & $0.62 \pm 0.14$ & $<0.67$ & $39.60 \pm 0.43$ & $<39.63$ & \dots \\
    J1007+1239 & $2.71\pm0.77$ & $0.74\pm0.21$ & $<1.25$ & $39.15 \pm 0.43$ & $<39.38$ & \dots \\
    J1024+1959 & $<2.56$ & $<0.70$ & $<1.10$ & $<40.57$ & $<40.76$ & \dots \\
    J1034+5803 & $106.65\pm11.56$ & $29.10\pm3.18$ & $14.30\pm3.58$ & $40.56 \pm 0.42$ & $40.25 \pm 0.43$ & $0.05 \pm 2.82$\\
    J1128+1716 & $23.81\pm2.18$ & $6.55\pm0.60$ & $<2.03$ & $40.26\pm0.42$ & $<39.75$ & \dots \\
    J1139+1934 & $<2.76$ & $<0.76$ & $<1.15$ & $<38.75$ & $<38.93$ & \dots \\
    J1158+2618 & $<2.04$ & $<0.56$ & $<1.46$ & $<39.83$ & $<40.24$ & \dots \\
    J1200+3151 & $<7.80$ & $<2.13$ & $<2.99$ & $<39.71$ & $<39.86$ & \dots \\
    J1213+3638 & $1025.02\pm11.72$ & $282.00\pm3.20$ & $218.12\pm4.13$ & $39.52 \pm 0.41$ & $39.41 \pm 0.42$ & $0.36 \pm 0.04$\\
    J1221+1135 & $10.94\pm0.73$ & $3.01\pm 0.20$ & $2.01\pm0.27$ & $38.81 \pm 0.42$ & $38.63 \pm 0.42$ & $0.25 \pm 0.37$\\
    J1221+1130 & $<7.12$ & $<1.96$ & $<2.57$ & $<37.45$ & $<37.56$ & \dots \\
    J1221+1130 & $127.10\pm3.81$ & $34.71\pm1.04$ & $11.60\pm0.92$ & $38.58 \pm 0.42$ & $38.10 \pm 0.42$ & \dots \\
    J1235+3931 & $6.14\pm0.95$ & $1.69\pm0.26$ & $<1.31$ & $40.22 \pm 0.43$ & $<40.11$ & \dots \\
    J1235+3931 & $318.80\pm3.56$ & $87.70\pm0.98$ & $43.31\pm1.03$ & $40.13 \pm 0.41$ & $39.83 \pm 0.42$ & $ -0.17 \pm 0.23$\\
    J1241+3324 & $2.67\pm0.62$ & $0.73\pm0.17$ & $<0.74$ & $39.94 \pm 0.43$ & $<39.94$ & \dots \\
    J1259+3450 & $55.28\pm6.19$ & $15.21\pm1.69$ & $5.36\pm1.74$ & $39.38 \pm 0.42$ & $38.93 \pm 0.43$ & $0.06 \pm 0.22$\\
    J1307+5426 & $7.66\pm0.88$ & $2.09\pm0.24$ & $1.14\pm0.32$ & $40.71 \pm 0.42$ & $40.45 \pm 0.43$ & $-0.14 \pm 1.10$\\
    J1308+1130 & $<5.68$ & $<1.55$ & $<2.71$ & $<40.34$ & $<40.58$ & \dots \\
    J1320+3308 & $4.07\pm0.92$ & $1.11\pm0.25$ & $1.56\pm0.33$ & $39.92 \pm 0.43$ & $40.06 \pm 0.43$ & $ 0.11 \pm 1.49$\\
    J1328+4355 & $6.69\pm1.16$ & $1.84\pm0.32$ & $<1.31$ & $40.52 \pm 0.43$ & $<40.22$ & \dots \\
    J1329+3234 & $10.83\pm0.91$ & $2.98\pm0.25$ & $2.45\pm0.36$ & $40.21 \pm 0.42$ & $40.13 \pm 0.43$ & $0.39 \pm 0.29$\\
    J1344+5601 & $4.79\pm1.05$ & $1.64\pm0.36$ & $<1.37$ & $41.30 \pm 0.43$ & $<41.22$ & \dots \\
    J1420+3954 & $5.31\pm1.10$ & $1.45\pm0.30$ & $<1.29$ & $39.00 \pm 0.43$ & $<38.95$ & \dots \\
    J1424+2254 & $<1.10$ & $<0.30$ & $<0.47$ & $<39.87$ & $<40.07$ & \dots \\
    J1548+3517 & $<2.69$ & $<0.74$ & $<1.28$ & $<40.73$ & $<40.97$ & \dots \\
    J1606+0809 & $<12.20$ & $<3.33$ & $<6.16$ & $<39.81$ & $<40.08$ & \dots \\
    J2128+0003 & $<2.91$ & $<0.80$ & $<1.37$ & $<40.25$ & $<40.48$ & \dots \\
    \noalign{\smallskip}
    \hline
        
    \end{tabular}
    
    {$^*$Hardness ratios, defined as (H-S)/(H+S), are only calculated for sources in which there are $3\sigma$ detections in both the soft (S; 0.3--2 keV) and hard (H; 2--10 keV)} bands.
    \label{sample_luminosities}
\end{table*}



\subsection{Spectroscopic Case Study}

In addition to our analysis of the X-ray fluxes from this sample, we also extracted spectra from all sources that had a 3$\sigma$ detection. Of those, only one (J121345.22+363805.8, hereafter J1213+3638) had sufficient signal-to-noise to perform robust spectral fitting. This source had $\approx$3500 counts in each of the MOS1 and MOS2 images. The pn data were heavily affected by background flares and therefore are not taken into account for this analysis. We grouped the spectrum at a minimum of 20 counts per bin in order to use $\chi^2$ statistics. The spectrum was background-subtracted and analyzed in \textsc{xspec} \citep{arnaud1996}, where we fit a simple absorbed power law (\textsc{phabs*zphabs*zpowerlw}) with the fixed \textsc{phabs} component corresponding to the Galactic $N_H = 2.46\times 10^{20}$~cm$^{-2}$ as reported by the Swift $N_H$ tool \citep{willingale2013}. The best-fit power law photon index was determined to be $\Gamma=1.92\pm0.06$, with an associated normalization of $(8.3\pm0.4) \times 10^{-4}$ photons~keV$^{-1}$~cm$^{-2}$~s$^{-1}$. While the target's hard X-ray luminosity of $\approx 10^{39.5}$~erg~s$^{-1}$ could be suggestive of a low luminosity AGN, this source was previously reported as a ULX powered by a stellar mass black hole \citep[e.g., ][]{ghosh2021}. The best-fit model returned an intrinsic $N_{\rm{H}}$ column density of $(2.26 \pm 0.26) \times10^{21}$~cm$^{-2}$. 

\begin{figure}[h]
\centering
\includegraphics[width=0.44\textwidth]{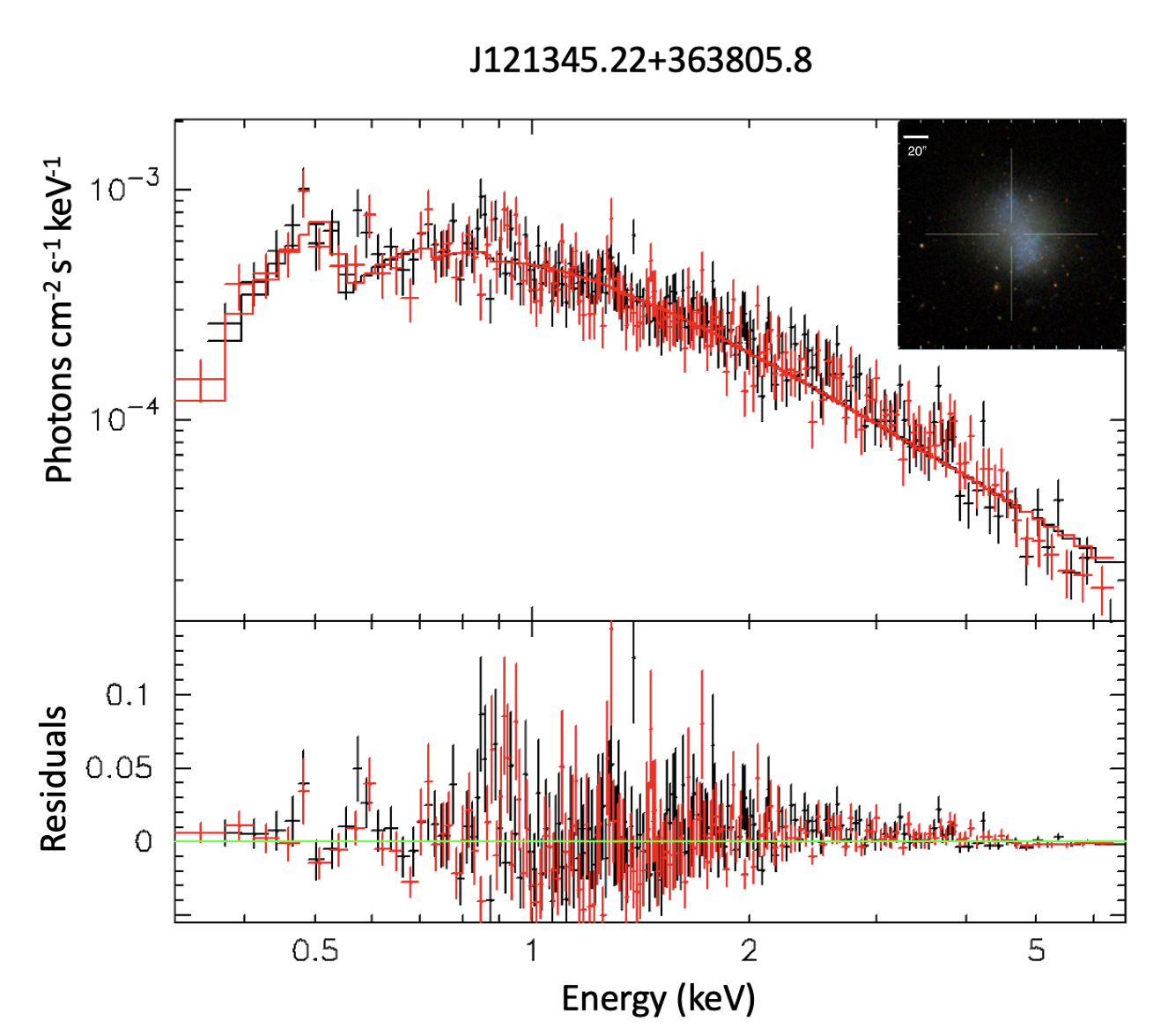} \\
\caption{\footnotesize 
Unfolded MOS1 (black) and MOS2 (red) spectra and residuals (data - folded model) for J1213+3638. The inset shows the SDSS thumbnail image of this target. As can be seen, this spectrum appears to be fit well by an absorbed power-law component.}
\label{specfit-final}
\end{figure}

\section{Discussion and Implications} 
\label{sec:implications}

\subsection{Multiwavelength Relationships}

In order to better understand the X-ray emission of low metallicity dwarf galaxies as a whole, we use several $L_{0.3-10\,\mathrm{keV}}$ and $L_{2-10\,\mathrm{keV}}$ relations to compare our target sample against the control sample. First, we explore the relation between mid-infrared luminosity and X-ray luminosity, as derived through the Python \textsc{linmix}\footnote{https://linmix.readthedocs.io/en/latest/src/linmix.html} package \citep{kelly2007}, which uses Bayesian inference and a Markov Chain Monte Carlo simulation to perform linear regressions on a sample of data, taking upper limits into account with a separate likelihood function (see Eq. 42 in \citet{kelly2007}). \textbf{The MCMC chains from this fitting procedure can be found in a Zenodo repository\footnote{doi:10.5281/zenodo.8320016}.} Mid-infrared luminosities are traced by the Wide-field Infrared Survey Explorer (WISE) W2 band ($4.3\,\mu$m), taken from the ALLWISE catalog \citep{cutri2013} with SNRs $>3$.

\begin{figure*}
\centering
\includegraphics[width=0.45\textwidth]{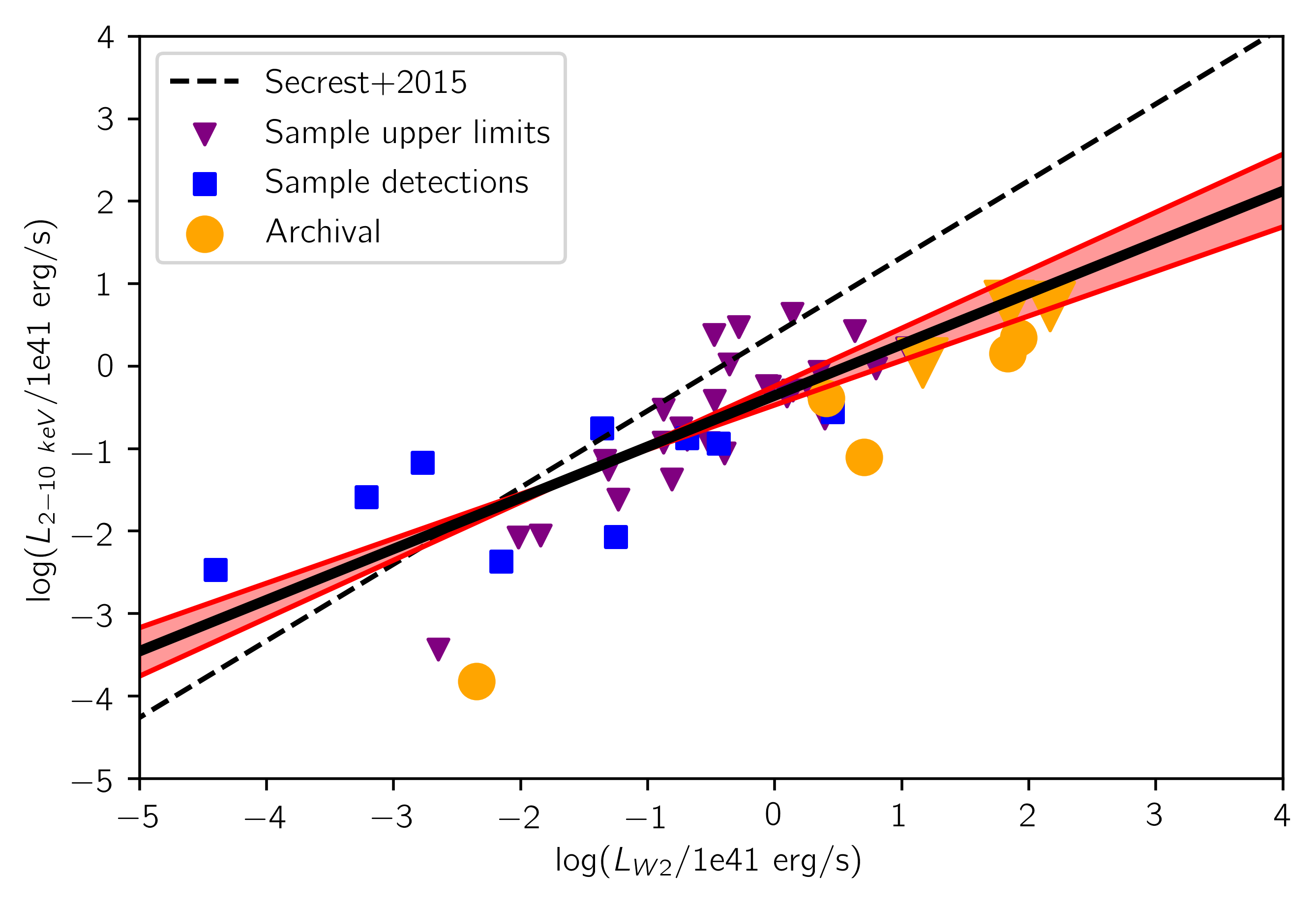} \includegraphics[width=0.35\textwidth]{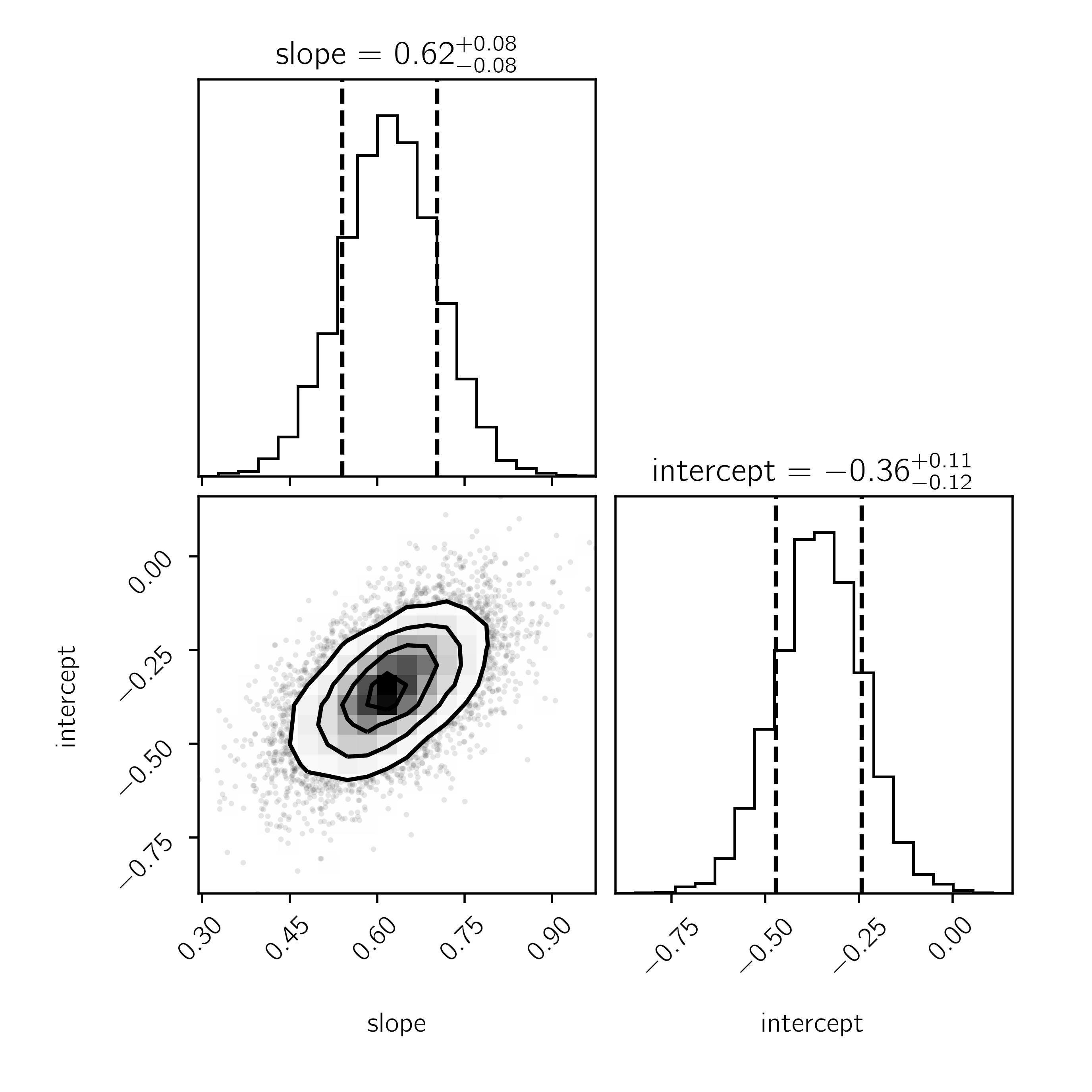}  \\
\includegraphics[width=0.45\textwidth]{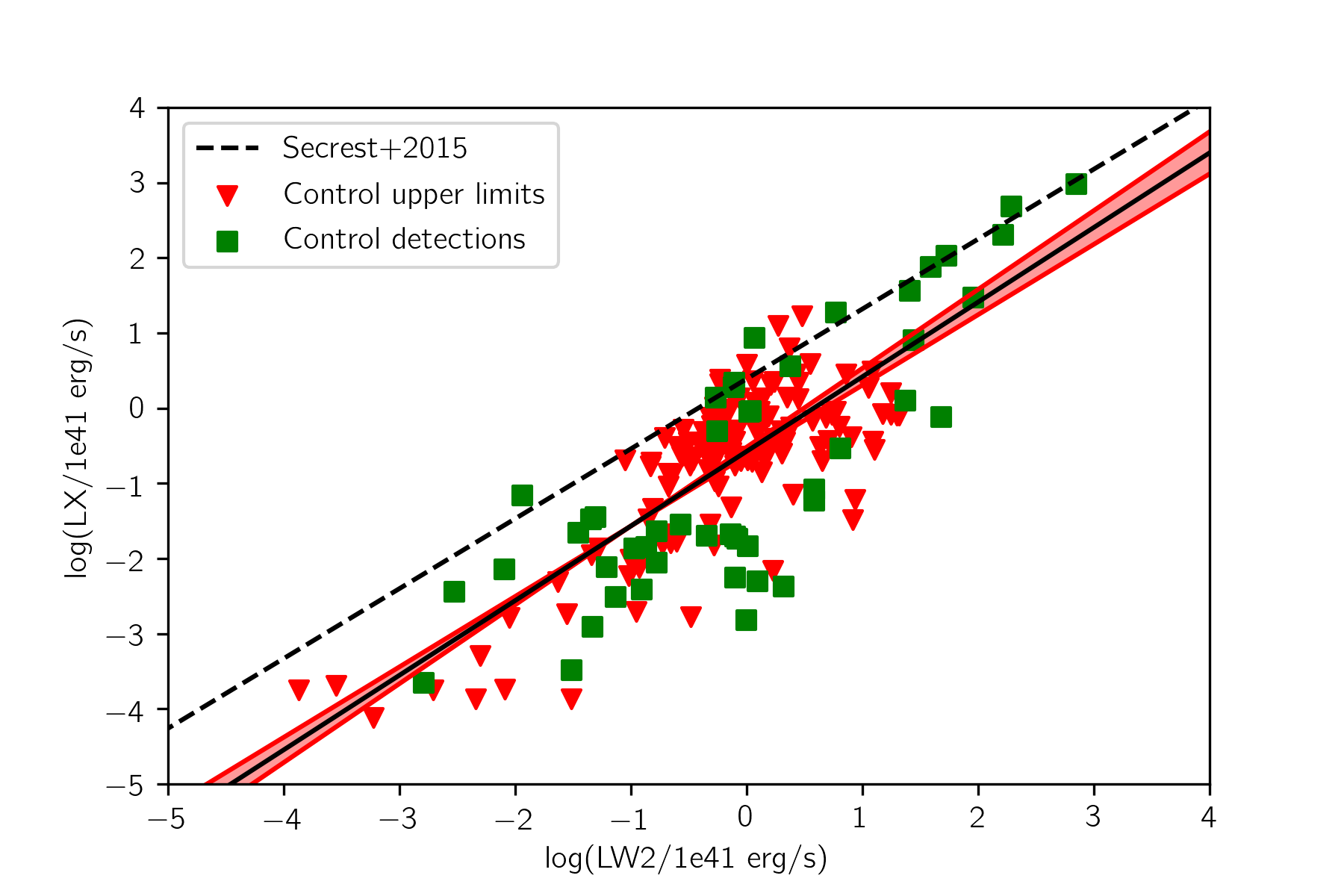}
\includegraphics[width=0.35\textwidth]{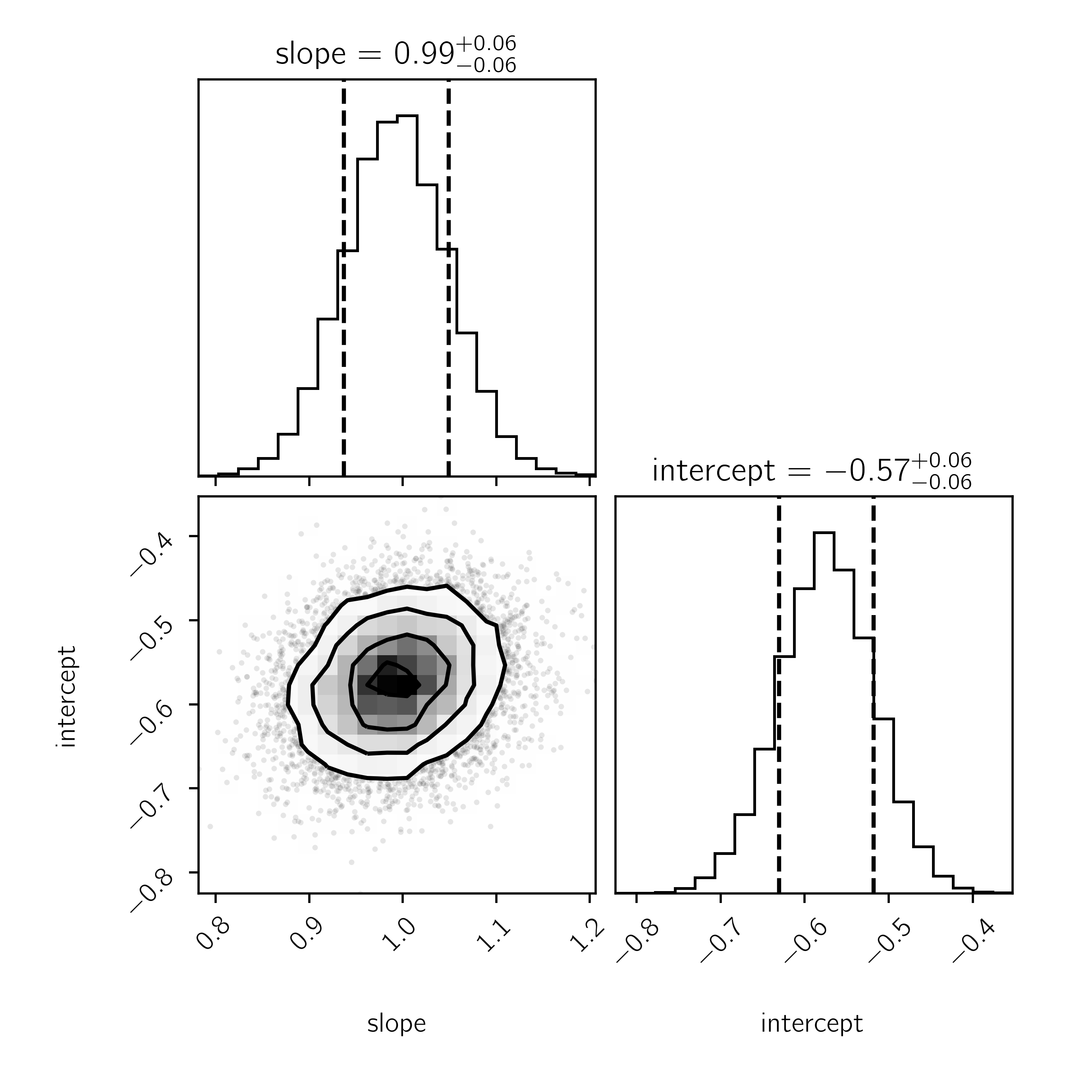}\\
\includegraphics[width=0.45\textwidth]{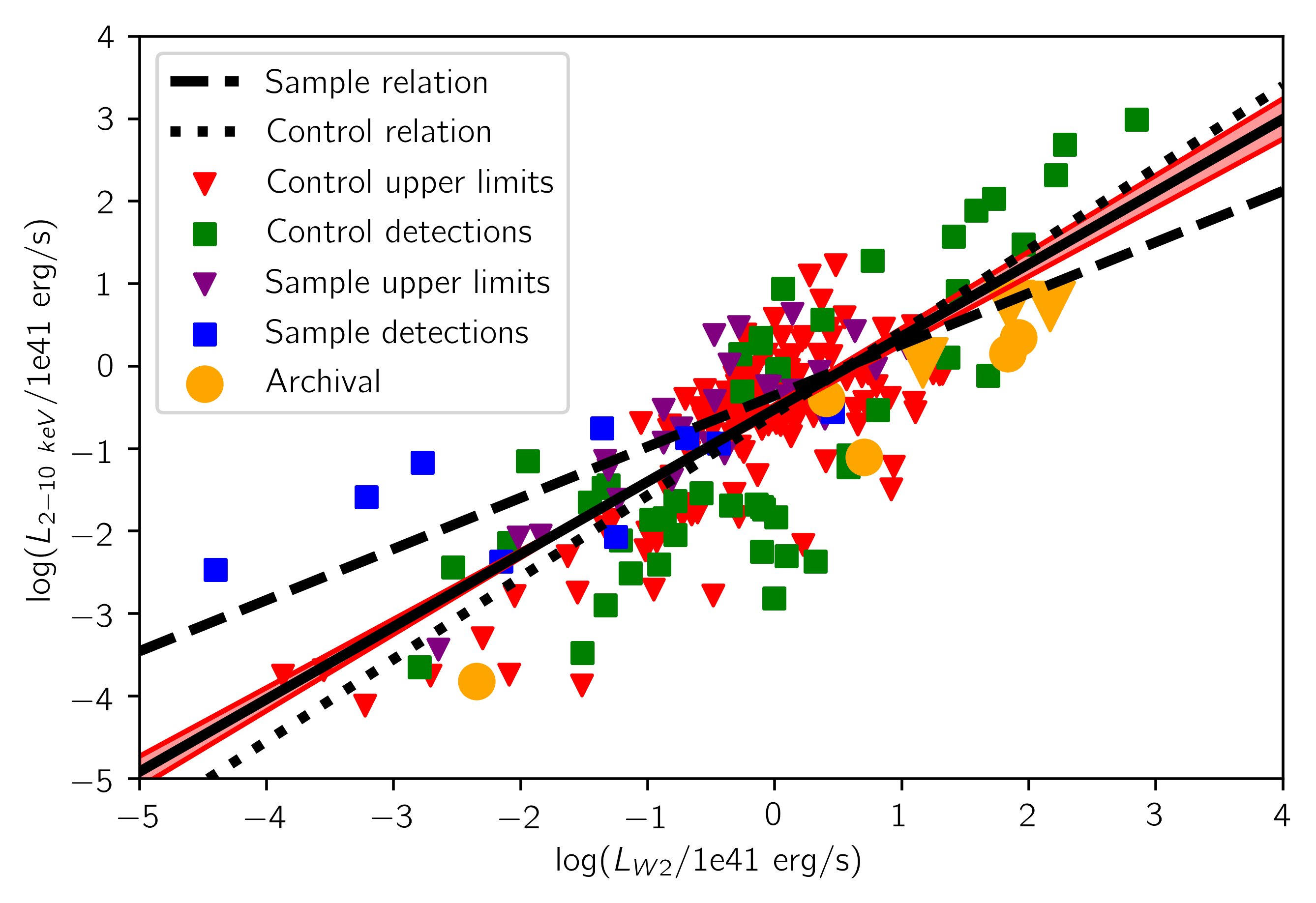}
\includegraphics[width=0.35\textwidth]{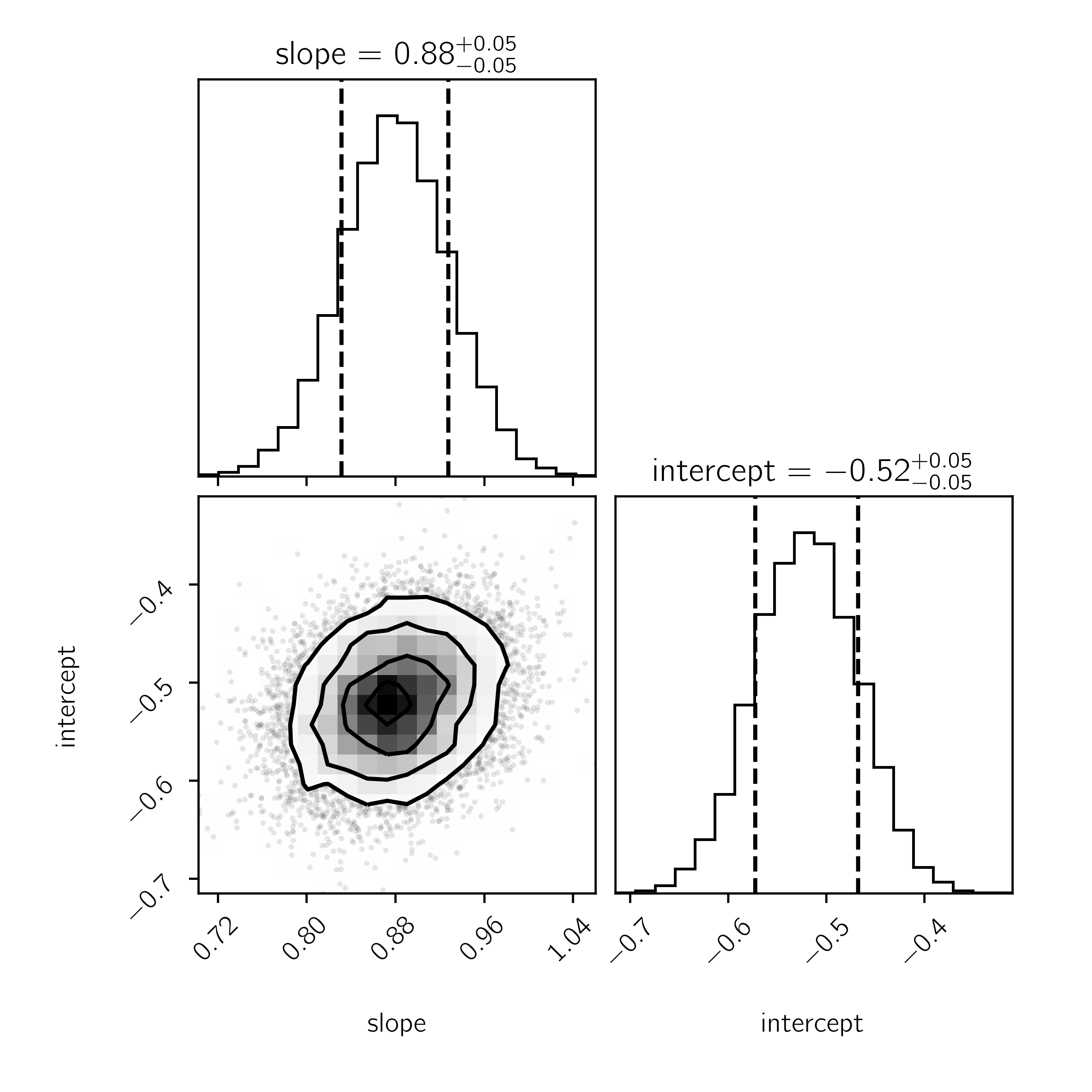}\\

\caption{\footnotesize Left: Relations between the W2 luminosity and the 2--10 keV X-ray luminosity for the low metallicity sample (top left), control sample (middle left), and full (low metallicity + control) sample (bottom left) in this study. Upper limits and detections are denoted by blue triangles and black squares, respectively. The low metallicity galaxies with apparent X-ray deficits reported in \citet{simmonds2016, cann2020} are plotted in orange. The \textsc{linmix}-derived relation and samples from the Bayesian posterier are plotted in black, with the 16\% and 84\% quartiles shaded in red. For the figures showing the sample and control relations, the relation reported in \citet{secrest2015} for bulgeless galaxies is plotted as a black dashed line. For the figure showing the full sample relation, the relations derived for the low metallicity and control samples are plotted in dashed and dotted lines, respectively. Right: The corner plot from the MCMC chains, showing the contours in parameter space and the distribution histograms for the low metallicity sample (top right), control (middle right), and full sample (bottom right), with +/-1$\sigma$ plotted as dashed lines.}
\label{lxlw2_sample}
\end{figure*}

As can be seen in Figure \ref{lxlw2_sample}, for a given mid-infrared luminosity, low metallicity galaxies have systematically higher X-ray luminosities than their higher metallicity counterparts, up until a turnover point of $L_{W2} \approx 10^{42}$~erg~s$^{-1}$. As the majority of these galaxies likely have their X-ray emission dominated by XRBs, it is likely that the source of this discrepancy is due to the increased luminosities of XRBs reported in low metallicity galaxies \citep[e.g.,][]{fragos2013,lehmer2019}. This enhancement results in higher X-ray luminosities that are not necessarily tied to a similar increase in W2 luminosity due to a relative lack of dusty nebulae surrounding these sources. This results in a relatively shallow relation between the two parameters compared to the control sample where those two are more directly correlated. To explore this we further compared the relation between $SFR$ and $L_{W2}$ of the two populations, and found that, while the low metallicity sample showed a fairly tight relation, the control sample showed significant upper spread, suggestive of additional non-SFR sources of W2 emission (e.g., AGNs) compared to the low metallicity population. 
We report the two relations below:\\

$\log(\frac{L_{2-10\,\rm{keV}}}{10^{41}~ \textrm{erg s}^{-1}}) = 0.62 \times \log(\frac{L_{W2}}{10^{41}~ \textrm{erg s}^{-1}}) - 0.36 $ \\

\noindent for ($\log$[N~II]/H$\alpha$) $< -1.0$ \textbf{(intrinsic scatter: 0.5828, reduced $\chi^2$: 0.72)},  and\\


$\log(\frac{L_{2-10\,\rm{keV}}}{10^{41}~ \textrm{erg s}^{-1}}) = 0.99 \times \log(\frac{L_{W2}}{10^{41}~ \textrm{erg s}^{-1}}) - 0.57 $ \\

\noindent for ($\log$[N~II]/H$\alpha$) $> -1.0$ \textbf{(intrinsic scatter: 0.7680, reduced $\chi^2$: 0.89)}.

For completeness, we also report the $L_{W2}-L_{2-10\,\rm{keV}}$ relation for the full (low metallicity + control) sample \textbf{(intrinsic scatter: 0.7701, reduced $\chi^2$: 1.2)}:\\


$\log(\frac{L_{2-10\,\rm{keV}}}{10^{41}~ \textrm{erg s}^{-1}}) = 0.88 \times \log(\frac{L_{W2}}{10^{41}~ \textrm{erg s}^{-1}}) - 0.52 $ \\

\textbf{The covariance matrices for the sample, control, and full relations, respectively, are:}

\begin{center}
$\begin{bmatrix}
0.0070 & 0.0054\\
0.0054 & 0.0137
\end{bmatrix}$\vspace{3mm}

$\begin{bmatrix}
0.0032 & 0.0006\\
0.0006 & 0.0032
\end{bmatrix}$\\\vspace{3mm}

$\begin{bmatrix}
0.0024 & 0.0006\\
0.0006 & 0.0027
\end{bmatrix}$\\
\end{center}

\textbf{Note that the correlation of the slope and intercept with the intrinsic scatter is negligible and therefore the intrinsic scatter is not included in this matrix.}

Because currently reported $L_X$ relations are typically calculated using high mass, high metallicity galaxies and well-known AGNs \citep[e.g.,][]{lamastra2009,dong2012,secrest2015}, this relation can provide insight for studies of dwarf galaxies regardless of metallicity. Given the ubiquity of archival WISE observations, the above relations may provide useful estimates for future X-ray observations targeting dwarf galaxies. In particular, we note that the low metallicity AGNs with reported X-ray deficits \citep{simmonds2016,cann2020} pictured in Figure \ref{lxlw2_sample} are mostly in reasonable agreement with any of the derived relations, suggesting that the larger driving force behind the purported X-ray deficit may instead be due to the relations calculated for higher mass galaxies being less applicable for dwarf galaxies.

\begin{figure}[h]
\centering
\includegraphics[width=0.44\textwidth]{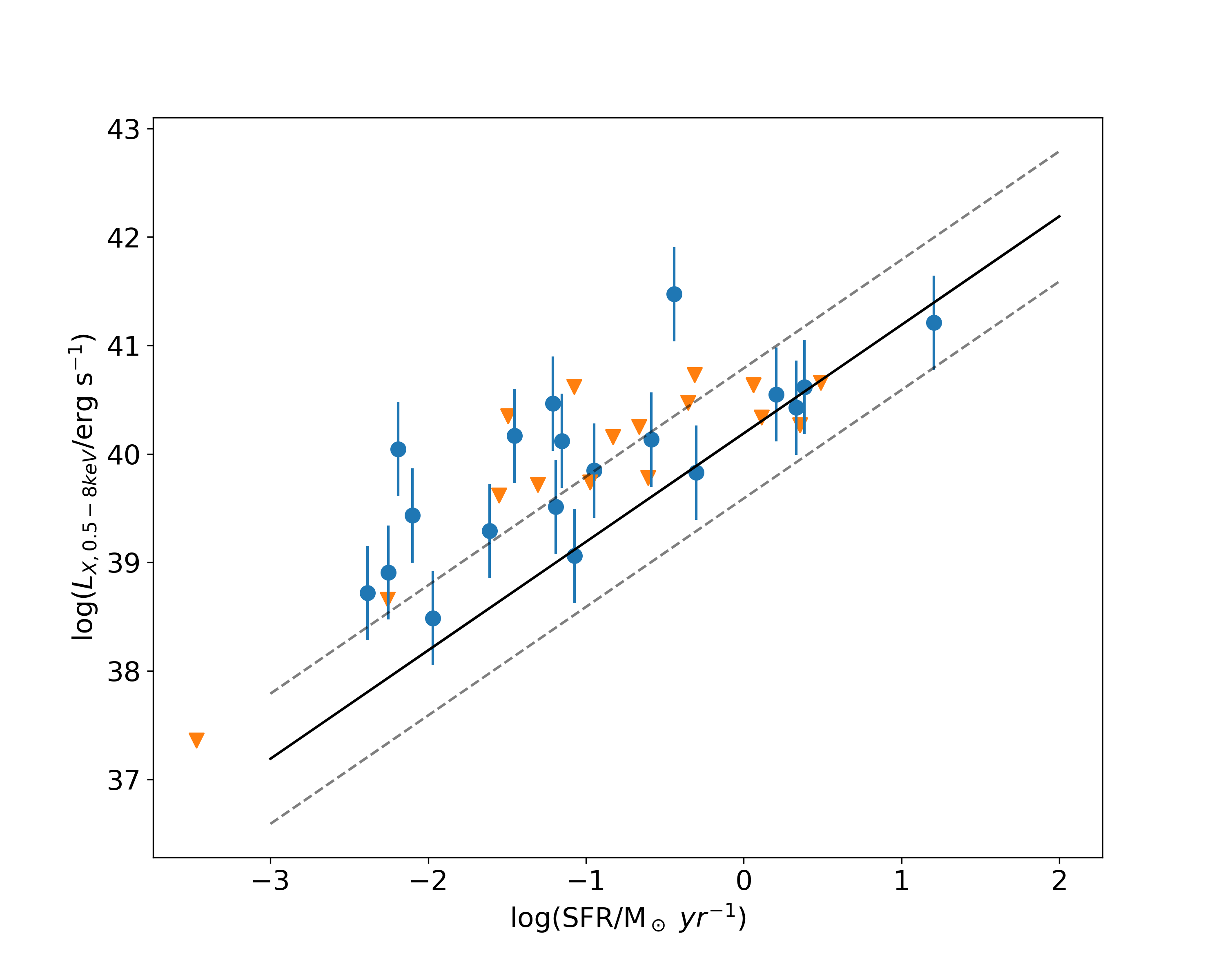}
\caption{\footnotesize The 0.5-8 keV X-ray luminosities and upper limits of the low metallicity dwarf sample plotted against the SFR-L$_X$ relation (black line) reported in \citet{lehmer2022}. Error bars are plotted for all detections.}
\label{lx_sfr}
\end{figure}

There have been multiple studies on the relation between the star-forming rate (SFR) and the X-ray luminosity in low metallicity galaxies \citep[e.g.,][]{fragos2013,lehmer2022} that attempt to take into consideration the enhanced emission from stellar processes. In Figure~\ref{lx_sfr}, we compare our sample to the recently reported $L_{0.5-8\,\mathrm{keV}}$--SFR relations in \citet{lehmer2019,lehmer2022}. Note that to compare with the relation presented in those works, we use PIMMS to convert our full band 0.3--10 keV luminosities to the Chandra 0.5--8 keV band. As can be seen in the figure, our sample deviates from the SFR-$L_{0.5-8\,\mathrm{keV}}$ relation presented in \citet{lehmer2019,lehmer2022}, however, based on the lack of sensitivity of the upper limits, this is likely due to our sample probing the high luminosity scatter from the relation.


\begin{figure}[h]
\centering
\includegraphics[width=0.99\linewidth]{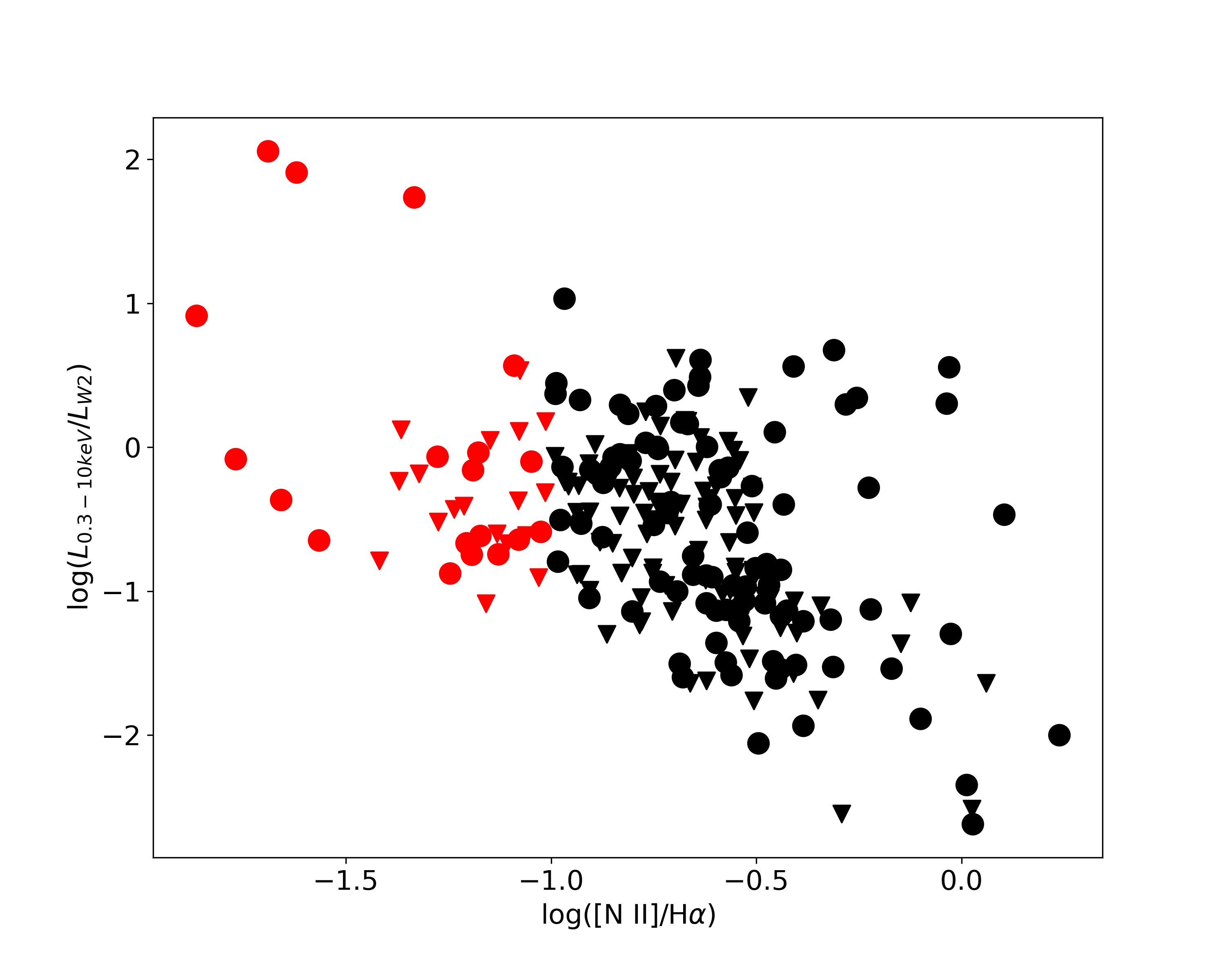} \\
\includegraphics[width=0.99\linewidth]{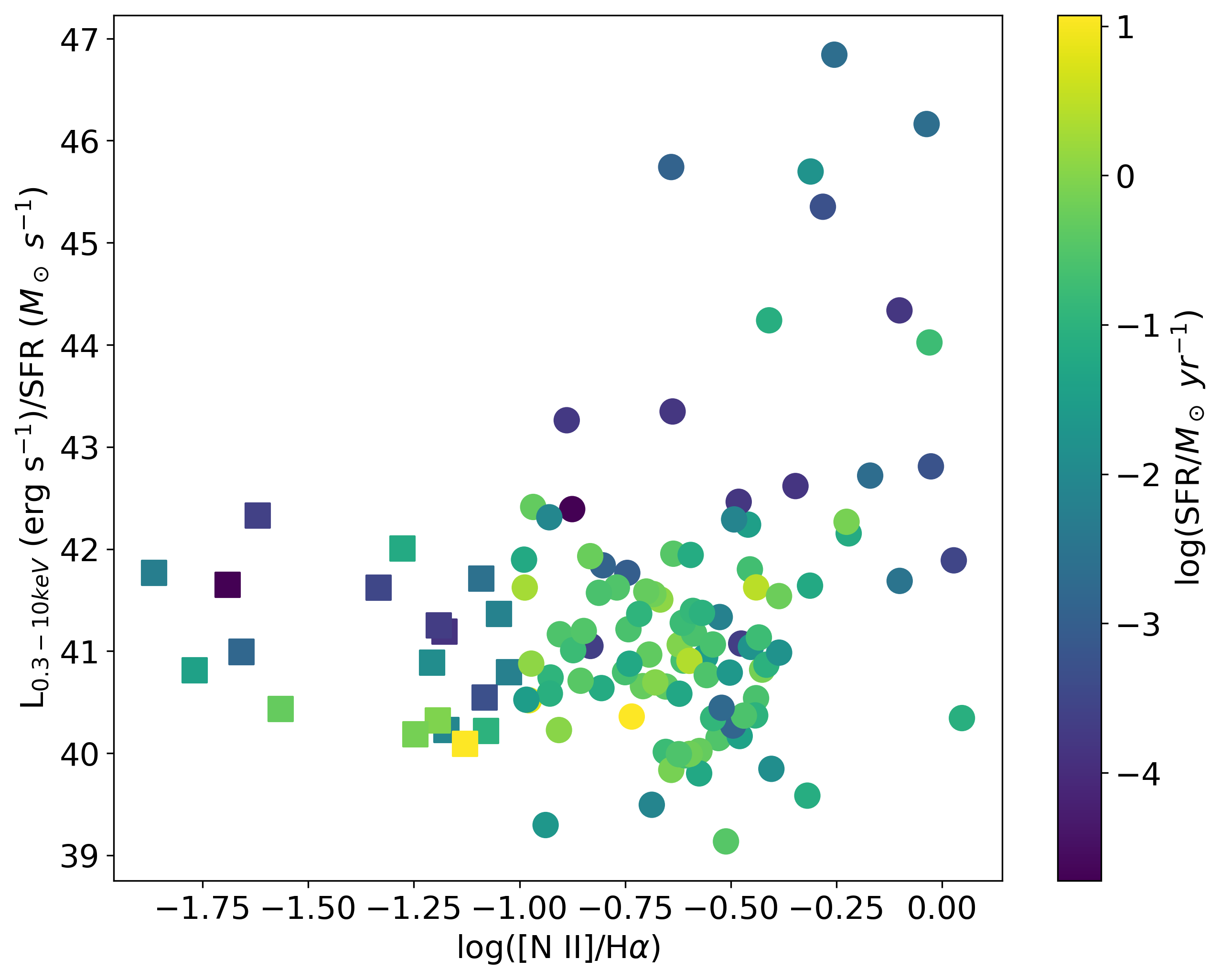} \\
\caption{\footnotesize
Top: The ratio of the full band X-ray luminosity detections (circles) and upper limits (triangles) over the W2 luminosity, as a function of log([\ion{N}{2}]/H$\alpha$) for the low metallicity sample (red) and the control sample (black). Bottom: The ratio of the full band X-ray luminosity over the star-forming rate for 3$\sigma$ detections from the low metallicity (squares) and control (circles) samples. The point color is representative of the SFR, shown by the colorbar.} 
\label{metal_relations}
\end{figure}

Finally, we explore the metallicity dependence of the X-ray luminosity, normalized by various parameters, including $L_X/L_{W2}$ and $L_X$/SFR, shown in Figure \ref{metal_relations}. Despite the many upper limits due to limited sensitivity in this archival work, an anti-correlation (Pearson coefficient $=-0.50$) can be seen, where $L_X/L_{W2}$ increases with decreasing metallicity.  This is in agreement with the relations presented in Figure \ref{lxlw2_sample}, in which low metallicity galaxies show systematically higher $W2$ luminosities for a given $L_{2-10\,\mathrm{keV}}$ when compared to their higher metallicity counterparts. 

Also pictured in Figure \ref{metal_relations} is the relation between metallicity and the ratio of $L_{0.3-10\,\mathrm{keV}}$/SFR. While there is no clear correlation with metallicity (Pearson coefficient $= 0.26$), it is interesting to see the difference in spread between the two populations. While the low metallicity sample has a fairly narrow range of values for $L_{0.3-10\,\mathrm{keV}}$/SFR, there is considerable spread seen for the control sample. As shown by the colorbar, the spread is typically comprised of sources with low SFRs, which are less present in the low metallicity sample detections. This could suggest that the X-ray luminosities from these control targets are from sources independent of star formation. This spread is seen most prominently in the highest metallicity end of the high metallicity sample, implying that enhanced X-ray emission from an AGN is more prevalent in that extreme. 
A larger, deeper study of low metallicity galaxies is needed  to determine if this trend continues, or if there may be similar spread with a larger sample size.

\subsection{Candidate AGNs}
\begin{figure}[h]
\centering
\includegraphics[width=0.99\linewidth]{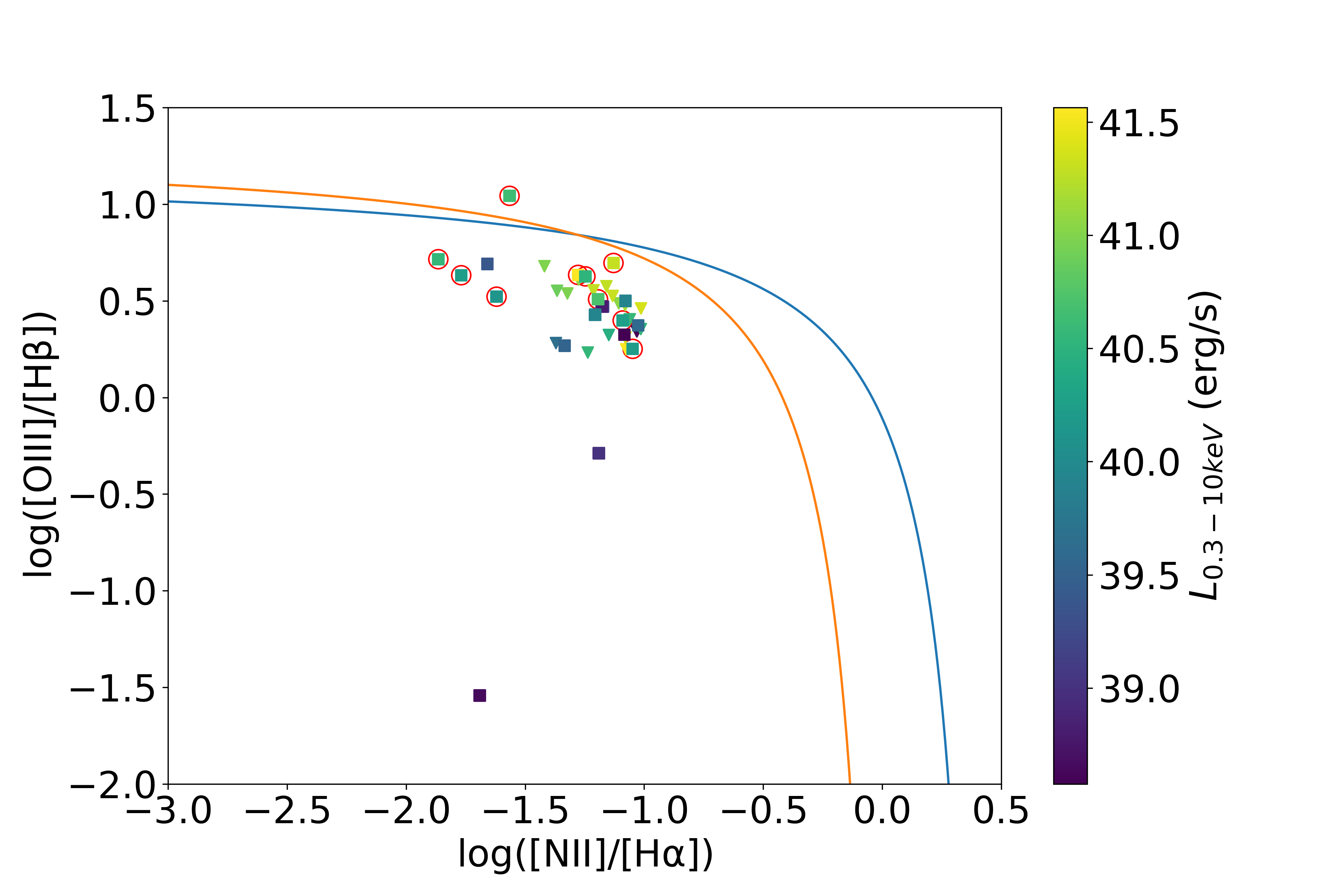} \\
\includegraphics[width=0.99\linewidth]{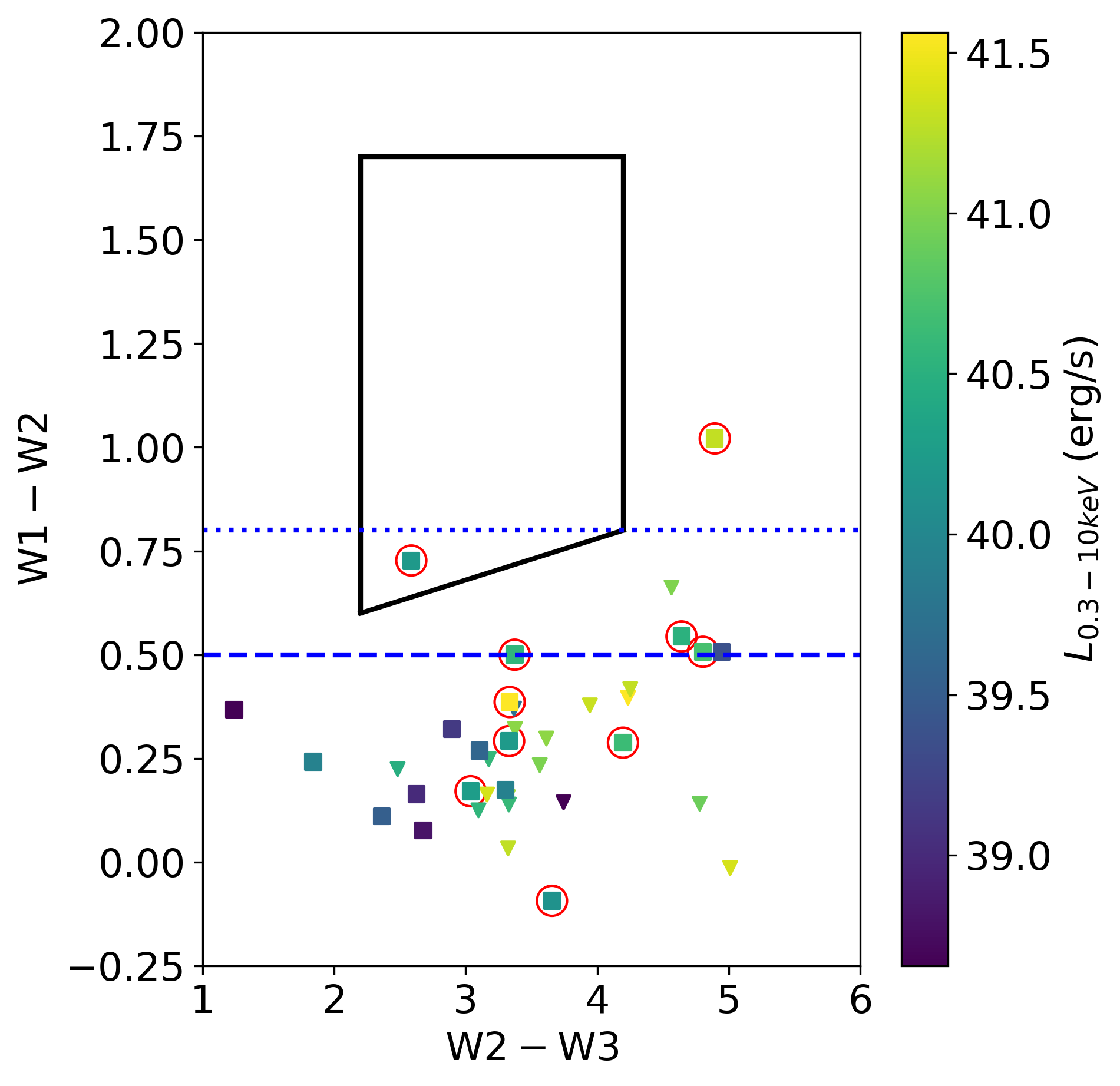}\\
\caption{\footnotesize
Top: BPT diagram displaying the emission line ratios of the low metallicity dwarf sample, with the AGN \citep{kewley2001} and star-forming \citep{kauffmann2003} delineations in blue and orange, respectively. Detected sources are plotted as squares, and upper limits are plotted as triangles. The color corresponds to the $\log$L$_{0.3-10\,\rm{keV}}$ of the detection or upper limit. Bottom: Mid-infrared color-color diagram with the low metallicity sample plotted similarly to that in the BPT diagram. The wedge corresponds to the \citet{jarrett2011} demarcation, and the two horizontal lines correspond to W1-W2 = 0.8 (dotted) and 0.5 (dashed) \citep{stern2012,blecha2018}. In both figures, sources that have $L_{X,\,0.3-10\,\rm{keV}} > 10^{40}$~erg~s$^{-1}$ are circled in red.} 
\label{bptjarrett}
\end{figure}

In this section, we take a deeper look at the ten targets that displayed full-band X-ray luminosities above $10^{40}$~erg~s$^{-1}$ (Table~\ref{sample_luminosities}). In Figure~\ref{bptjarrett}, we plot the low metallicity dwarf sample on the Baldwin-Philips-Terlevich (BPT) optical emission line ratio \citep{baldwin1981,kewley2001,kauffmann2003} and WISE mid-infrared color-color \citep{jarrett2011,stern2012} diagrams used to identify galaxy emission as either primarily star-forming or AGN in origin. In these plots, the color of the points corresponds to the detected or upper limit X-ray luminosity of the galaxies as shown in the color bar. Galaxies with X-ray luminosities above $10^{40}$~erg~s$^{-1}$ are circled in red. As can be seen, the majority of our sources, whether detected or not, have optical emission suggestive of primarily star formation. Indeed, the majority of detected sources have luminosities comparable to those expected from stellar activity, however it does not necessarily explain the higher luminosity sources. This is not unexpected, as the use of the [\ion{N}{2}]/H$\alpha$ metallicity diagnostic typically selects few BPT-selected AGNs. In fact, in a study of $\approx23,000$ Seyfert 2 galaxies, only 40 were identified as low metallicity AGN using the [\ion{N}{2}]/H$\alpha$ and BPT diagnostics \citep{groves2006}.

Comparing these multi-wavelength diagnostics against one another for the detected X-ray sources, of the ten sources with X-ray luminosities $>10^{40}$~erg~s$^{-1}$, only one (4\%) candidate AGN in this sample was identified through the optical BPT emission line diagnostics (compared to 13\% in the control sample) and only one (4\%) \citep{jarrett2011,stern2012} or five (16\%) others \citep{blecha2018} were found using mid-infrared diagnostics, depending on the selection criterion used (compared with 4\% and 8\% for the control sample). We further explore the $L_U/L_{0.3-10 keV}$ ratio for these targets, and find that one of the AGN candidates has a low ($\log(L_U/L_X) \approx 0.6$) ratio, further suggestive of AGN activity. The remaining candidates did not show ratios particularly indicative of an AGN. Based on these discrepancies, the AGN candidates unearthed from this study may probe a different population than those identified from these commonly used optical and mid-infrared diagnostics. Further, none of these targets were reported in \citet{reefe2022a} to have optical coronal lines detected by SDSS, nor do they show evidence of broad spectroscopic emission lines. The lack of overlap in AGN candidate detection across these diagnostics demonstrates a potentially significant bias in the use of these tools in searching for AGNs in this low mass, low metallicity population. As the development of these diagnostics typically involved solely high mass, high metallicity AGNs, it is not unreasonable that these same diagnostics will be ineffective in unearthing AGNs in other populations. Further study, particularly involving high sensitivity, multi-wavelength observations of a larger sample of \textit{confirmed} AGNs in these environments, is needed to determine if there are other multi-wavelength diagnostics that can effectively identify candidate AGNs in the low mass and low metallicity regime.


We also explore the potential for this emission to have been produced by XRBs. In Figure \ref{lx_sfr}, we plot the derived 0.5-8 keV luminosities as a function of SFR along with the relation from \citet{lehmer2022} to show the expected X-ray contribution from XRBs. As can be seen, a number of our detections fall well above the expected emission from XRBs, including three with $L_X > 10^{40}$~erg~s$^{-1}$, providing more supporting evidence for the presence of candidate AGNs in these sources. 



\subsection{Comparison with Previous Studies}

Here we discuss the findings of this paper in connection to previous studies of dwarf and low metallicity galaxies. As mentioned above, this work was partially motivated to explore how common the apparent X-ray deficit seen in the low metallicity dwarf galaxies reported in \citet{simmonds2016,cann2020} is in a wider sample. The relations used to determine the expected X-ray luminosity in these papers \citep{panessa2006, secrest2015} were derived for generally higher mass, higher metallicity galaxies, so there was a question as to whether new relations would be needed specifically for the low mass, low metallicity case. As shown in Section 4.1, this does seem to be true for the $L_{W2}-L_{2-10\, \mathrm{keV}}$ relation, as the one reported in \citet{secrest2015} is shown to be a poor fit for both the low metallicity and control samples, suggesting the need for a similar relation specific to dwarf galaxies in general at minimum. We therefore propose the above relations for use in this regime.

\citet{simmonds2016} cited an X-ray deficit in several of their sources as compared to the $L_{\mathrm{H\alpha}}-L_{2-10 \mathrm{keV}}$ and $L_{\mathrm{[O III]}}-L_{2-10 keV}$ relations reported in \citet{panessa2006}. With the larger sample in this work, however, we do not see a similar poor fit to either the low metallicity or control sample, suggesting that the apparent X-ray deficit in \citet{simmonds2016} is likely due to unique conditions within those particular sources.

We can further compare with the \citet{zou2023} study of dwarf galaxies in the XMM-SERVS field. Their sample shows a relatively small distribution of X-ray luminosities ($L_X \approx 10^{40-42}$ erg~s$^{-1}$) for their lowest-mass targets ($M_* \approx10^7$ M$_\odot$), with much greater variation ($L_X \approx 10^{40-44}$ erg~s$^{-1}$) in higher masses \citep[$M_* \approx10^9$ M$_\odot$; see Figure 6 in ][]{zou2023}. A similar analysis of our sample shows inconsistent results, with luminosities ranging from $L_X \approx 10^{38-44}$ regardless of galaxy mass for both the low metallicity and control samples. We note that the XMM-SERVS survey used in  \citet{zou2023} has significant improvements in sensitivity over a large subset of the targets reported here, and that the work probes a larger range of redshifts ($z < 1$).

Finally, we compare the results presented here to those in \citet{lehmer2021,lehmer2022}. As shown in Figure \ref{lx_sfr}, our low metallicity sample shows agreement with the upper scatter of the SFR-$L_{0.5-8 keV}$ relation in \citet{lehmer2022} for a similar sample of low metallicity galaxies. Similarly, when comparing our low metallicity sources with the metallicity vs. $L_{0.5-8\,\rm{keV}}/SFR$ relation reported in \citet{lehmer2021}, our sources tend to be consistent with the upper scatter in the relation. As this study primarily uses data from serendipitous observations, this is potentially due to a lack of sensitivity to sufficiently probe the lower scatter limits.

\subsection{Additional Considerations and Caveats}

The goal of this paper is to explore the X-ray properties of galaxies identified as low metallicity, and to compare them to their high metallicity counterpart. We emphasize that we have had to make reasonable assumptions in order to create a robust target and control sample.
First, our choice of [\ion{N}{2}]/H$\alpha$ for our metallicity diagnostic was informed by its applicability to both AGNs \citep[e.g.,][]{groves2006} and star-forming \citep[e.g.,][]{pettini2004} galaxies. While there are numerous metallicity estimates \citep[e.g.,][]{storchibergmann1998,tremonti2004,pilyugin2005,izotov2006,castro2017} that use more emission lines to reduce degeneracies, these assume a spectral energy distribution (SED) that is either primarily AGN or stellar in origin. For example, while the MPA/JHU catalog reports metallicity estimates, these estimates use the \citet{tremonti2004} diagnostic and are only calculated for star-forming galaxies. Furthermore, these diagnostics often require the use of the [\ion{O}{2}]$\lambda3726$, which is out of SDSS's wavelength coverage for $\approx40\%$ of our sample. To test the potential effects of different metallicity diagnostics on our results, we compared our derived metallicities with those reported by the MPA/JHU catalog \citep{tremonti2004}. We find that the estimates agree, with $\approx 90\%$ of the estimates for each sample being typically within $<0.5$ dex, and all targets retain their classification of ``low" or ``high" metallicity regardless of the diagnostic used. While there is still reasonable uncertainty in precise metallicity for these targets, the relative agreement between diagnostics and the consistent classification of all galaxies in either sample suggests that this source of uncertainty likely does not play a significant role in the interpretation of these results.

We also emphasize that, due to the lack of selection criteria for activity, we are in some cases comparing AGN to star-forming galaxies. For example, the high metallicity control sample has disproportionately more optically identified AGNs \citep{kewley2001,kauffmann2003} when compared to the low metallicity sample ($\approx 13\%$ vs. $\approx 4\%$). We note however, that there is an increasing number of candidate AGNs identified in low metallicity galaxies without optical evidence of an AGN \citep{cann2021,molina2021,reefe2022a}, and therefore it is probable that the true fraction of AGNs may be more consistent than initially evident. In fact, when comparing mid-infrared colors, the two samples show similar rates of mid-infrared colors suggestive of AGN activity whether by a 3-band \citep[][; $\approx 5\%$ of each sample]{jarrett2011} or a 2-band \citep[$\approx 5\%$ and $\approx10-15\%$ for][respectively]{stern2012,blecha2018} color cut. Overall, both samples show an approximately $20\%$ rate of AGN activity as identified by either optical or mid-infrared diagnostics, suggesting that the rate of AGN activity is likely similar in both populations, and therefore is not likely a source of significant uncertainty in the reported results.

Finally, we acknowledge that the sample reported here represents only a small fraction of both populations of dwarf galaxies. In particular, $<1\%$ of the full sample of low metallicity dwarf galaxies has been observed with XMM. While the reported sources can be considered a random sample due to the serendipitous nature of their observations, we emphasize that a more robust survey is needed to further determine if the trends reported in this study hold when exploring a larger sample. In the next section, we highlight future prospects to address this need.


\subsection{Future Prospects}
The results presented here clearly show the diversity of environments within the low metallicity dwarf galaxy population and the richness of information that can be gleaned by detailed, multi-wavelength observations. In this section, we address the implications of these results for current and future X-ray observatories. The use of XMM-Newton for this study was not only motivated by its large field of view for serendipitous observations, but also by its superior sensitivity. As many of these sources are faint in the X-rays, adequate sensitivity is necessary to robustly detect the emission from these objects. In potential future studies, XMM's spectral sensitivity can also be utilized to more completely study the emission structure from this population of galaxies, to create a more complete census of the X-ray emitting sources in this parameter space, and to determine if the spectrum discussed in this study is common.

With the all-sky survey from eROSITA \citep{salvato2022}, we have public access to sensitive X-ray imaging in the $0.2-2.3$ (soft) and $2.3-5.0$ (hard) keV bands for half of the sky.  As the limiting flux for the soft band is $\approx6.5 \times 10^{-15}$~erg~cm$^{-2}$~s$^{-1}$ \citep{salvato2022}, comparable to the fluxes detected in this study, the data made available from this mission will provide a valuable resource to explore the results presented here on a larger scale. In fact, in an analysis of the AGN fraction in dwarf galaxies in eFEDS, it was expected that the all-sky survey could uncover up to 1350 AGNs \citep{latimer2021}. The additional data available through this survey will allow for the analysis of a more robust sample than that presented in this work, and therefore will provide important insight to if the trends suggested by this study hold for a larger population.

There are also connections from this work to the upcoming missions, including Athena \citep{nandra2013} and the proposed NASA X-ray Probes. The current scope of Athena promises unparalleled imaging and spectroscopic sensitivity that will prove invaluable for deeper studies of this population, particularly those with dimmer X-ray luminosities than can be detected with current instrumentation. With the use of probe mission concepts such as the Line Emission Mapper \citep[LEM;][]{kraft2022}, high spectral resolution studies will provide detailed information on the soft emission line spectra of this galaxy population, including precise measurements of abundances of the hot gas. Further, as discussed in the introduction, low metallicity galaxies are sources of interest in the search for IMBHs, but the X-ray luminosities for these objects may be low and indistinguishable from stellar processes. LEM's superior spectral resolution will be invaluable to identifying emission lines from photoionization from an AGN, making it possible to disentangle stellar and IMBH-powered AGN contributions to the X-ray emission (Weaver et al., in prep).

\section{Conclusions}

We present an exploration of the X-ray properties of low metallicity dwarf galaxies. Our main results can be summarized as follows:

\begin{enumerate}

\item We note an excess in X-ray emission in a sample of low metallicity galaxies compared to those with higher metallicities.

\item We report $L_{W2}-L_{2-10 keV}$ relations for the low metallicity, control, and full samples for use in future X-ray studies of dwarf galaxies.


\item We highlight the detection of ten potential AGN candidates with 0.3-10 keV luminosities in excess of 10$^{40}$ erg~s$^{-1}$ in the low metallicity sample. The majority of these targets are not identified as AGNs through optical narrow emission line diagnostics, suggesting that X-rays may identify a sample that is hidden by traditional BPT diagnostics.
    
\end{enumerate}

The results presented here prove that there is a great diversity of X-ray emitting sources in low metallicity dwarf galaxies in terms of luminosity, spectral features, and multi-wavelength characterization (e.g. WISE colors), and have opened several questions about the prevalence of extreme stellar and AGN activity in this population. With more detailed spectral analyses of a larger sample with XMM, and with deeper, higher spectral resolution data from future missions, we will be able to study these galaxies in even greater detail and enhance our understanding of these local analogs of the earliest galaxies.

\begin{acknowledgments}
J.M.C.'s and R.W.P's research was supported by an appointment to the NASA Postdoctoral Program at the NASA Goddard Space, administered by Oak Ridge Associated Universities under contract with NASA. J.M.C. also acknowledges support from the XMM grant 80NSSC18K0758, and this publication makes use of data from XMM program 080324. This publication makes use of XMM data accessed through XMM Science Archive, provided by the European Space Agency. This research has made use of data and software provided by the High Energy Astrophysics Science Archive Research Center (HEASARC), which is a service of the Astrophysics Science Division at NASA/GSFC. This publication makes use of data products from the Wide-field Infrared Survey Explorer, which is a joint project of the University of California, Los Angeles, and the Jet Propulsion Laboratory/California Institute of Technology, funded by the National Aeronautics and Space Administration. This publication has also made use of data from SDSS. Funding for the Sloan Digital Sky Survey IV has been provided by the Alfred P. Sloan Foundation, the U.S. Department of Energy Office of Science, and the Participating Institutions. SDSS-IV acknowledges support and resources from the Center for High Performance Computing  at the University of Utah. The SDSS website is www.sdss.org.

SDSS-IV is managed by the Astrophysical Research Consortium for the Participating Institutions of the SDSS Collaboration including the Brazilian Participation Group, the Carnegie Institution for Science, Carnegie Mellon University, Center for Astrophysics | Harvard \& Smithsonian, the Chilean Participation Group, the French Participation Group, Instituto de Astrof\'isica de Canarias, The Johns Hopkins University, Kavli Institute for the Physics and Mathematics of the Universe (IPMU) / University of Tokyo, the Korean Participation Group, Lawrence Berkeley National Laboratory, Leibniz Institut f\"ur Astrophysik Potsdam (AIP),  Max-Planck-Institut f\"ur Astronomie (MPIA Heidelberg), Max-Planck-Institut f\"ur Astrophysik (MPA Garching), Max-Planck-Institut f\"ur Extraterrestrische Physik (MPE), National Astronomical Observatories of China, New Mexico State University, New York University, University of Notre Dame, Observat\'ario Nacional / MCTI, The Ohio State University, Pennsylvania State University, Shanghai Astronomical Observatory, United Kingdom Participation Group, Universidad Nacional Aut\'onoma de M\'exico, University of Arizona, University of Colorado Boulder, University of Oxford, University of Portsmouth, University of Utah, University of Virginia, University of Washington, University of Wisconsin, Vanderbilt University, and Yale University.

We also gratefully acknowledge the use of the software TOPCAT \citep{topcat} and Astropy \citep{astropy2018}.
\vspace{10.0cm}

\end{acknowledgments}

%

\vspace{5mm}
\facilities{XMM-Newton, WISE, SDSS}


\software{astropy \citep{astropy2013,astropy2018},  
          SAS \citep{sas}}

\bibliography{sample631}{}

\begin{thebibliography}{}
\expandafter\ifx\csname natexlab\endcsname\relax\def\natexlab#1{#1}\fi
\providecommand{\url}[1]{\href{#1}{#1}}
\providecommand{\dodoi}[1]{doi:~\href{http://doi.org/#1}{\nolinkurl{#1}}}
\providecommand{\doeprint}[1]{\href{http://ascl.net/#1}{\nolinkurl{http://ascl.net/#1}}}
\providecommand{\doarXiv}[1]{\href{https://arxiv.org/abs/#1}{\nolinkurl{https://arxiv.org/abs/#1}}}

\bibitem[{{Arnaud}(1996)}]{arnaud1996}
{Arnaud}, K.~A. 1996, in Astronomical Society of the Pacific Conference Series,
  Vol. 101, Astronomical Data Analysis Software and Systems V, ed. G.~H.
  {Jacoby} \& J.~{Barnes}, 17

\bibitem[{{Astropy Collaboration} {et~al.}(2013){Astropy Collaboration},
  {Robitaille}, {Tollerud}, {Greenfield}, {Droettboom}, {Bray}, {Aldcroft},
  {Davis}, {Ginsburg}, {Price-Whelan}, {Kerzendorf}, {Conley}, {Crighton},
  {Barbary}, {Muna}, {Ferguson}, {Grollier}, {Parikh}, {Nair}, {Unther},
  {Deil}, {Woillez}, {Conseil}, {Kramer}, {Turner}, {Singer}, {Fox}, {Weaver},
  {Zabalza}, {Edwards}, {Azalee Bostroem}, {Burke}, {Casey}, {Crawford},
  {Dencheva}, {Ely}, {Jenness}, {Labrie}, {Lim}, {Pierfederici}, {Pontzen},
  {Ptak}, {Refsdal}, {Servillat}, \& {Streicher}}]{astropy2013}
{Astropy Collaboration}, {Robitaille}, T.~P., {Tollerud}, E.~J., {et~al.} 2013,
  \aap, 558, A33, \dodoi{10.1051/0004-6361/201322068}

\bibitem[{{Astropy Collaboration} {et~al.}(2018){Astropy Collaboration},
  {Price-Whelan}, {Sip{\H{o}}cz}, {G{\"u}nther}, {Lim}, {Crawford}, {Conseil},
  {Shupe}, {Craig}, {Dencheva}, {Ginsburg}, {VanderPlas}, {Bradley},
  {P{\'e}rez-Su{\'a}rez}, {de Val-Borro}, {Aldcroft}, {Cruz}, {Robitaille},
  {Tollerud}, {Ardelean}, {Babej}, {Bach}, {Bachetti}, {Bakanov}, {Bamford},
  {Barentsen}, {Barmby}, {Baumbach}, {Berry}, {Biscani}, {Boquien}, {Bostroem},
  {Bouma}, {Brammer}, {Bray}, {Breytenbach}, {Buddelmeijer}, {Burke},
  {Calderone}, {Cano Rodr{\'\i}guez}, {Cara}, {Cardoso}, {Cheedella}, {Copin},
  {Corrales}, {Crichton}, {D'Avella}, {Deil}, {Depagne}, {Dietrich}, {Donath},
  {Droettboom}, {Earl}, {Erben}, {Fabbro}, {Ferreira}, {Finethy}, {Fox},
  {Garrison}, {Gibbons}, {Goldstein}, {Gommers}, {Greco}, {Greenfield},
  {Groener}, {Grollier}, {Hagen}, {Hirst}, {Homeier}, {Horton}, {Hosseinzadeh},
  {Hu}, {Hunkeler}, {Ivezi{\'c}}, {Jain}, {Jenness}, {Kanarek}, {Kendrew},
  {Kern}, {Kerzendorf}, {Khvalko}, {King}, {Kirkby}, {Kulkarni}, {Kumar},
  {Lee}, {Lenz}, {Littlefair}, {Ma}, {Macleod}, {Mastropietro}, {McCully},
  {Montagnac}, {Morris}, {Mueller}, {Mumford}, {Muna}, {Murphy}, {Nelson},
  {Nguyen}, {Ninan}, {N{\"o}the}, {Ogaz}, {Oh}, {Parejko}, {Parley}, {Pascual},
  {Patil}, {Patil}, {Plunkett}, {Prochaska}, {Rastogi}, {Reddy Janga},
  {Sabater}, {Sakurikar}, {Seifert}, {Sherbert}, {Sherwood-Taylor}, {Shih},
  {Sick}, {Silbiger}, {Singanamalla}, {Singer}, {Sladen}, {Sooley},
  {Sornarajah}, {Streicher}, {Teuben}, {Thomas}, {Tremblay}, {Turner},
  {Terr{\'o}n}, {van Kerkwijk}, {de la Vega}, {Watkins}, {Weaver}, {Whitmore},
  {Woillez}, {Zabalza}, \& {Astropy Contributors}}]{astropy2018}
{Astropy Collaboration}, {Price-Whelan}, A.~M., {Sip{\H{o}}cz}, B.~M., {et~al.}
  2018, \aj, 156, 123, \dodoi{10.3847/1538-3881/aabc4f}

\bibitem[{{Baldwin} {et~al.}(1981){Baldwin}, {Phillips}, \&
  {Terlevich}}]{baldwin1981}
{Baldwin}, J.~A., {Phillips}, M.~M., \& {Terlevich}, R. 1981, \pasp, 93, 5,
  \dodoi{10.1086/130766}

\bibitem[{{Barnes} \& {Hernquist}(1996)}]{barnes1996}
{Barnes}, J.~E., \& {Hernquist}, L. 1996, \apj, 471, 115,
  \dodoi{10.1086/177957}

\bibitem[{{Blecha} {et~al.}(2018){Blecha}, {Snyder}, {Satyapal}, \&
  {Ellison}}]{blecha2018}
{Blecha}, L., {Snyder}, G.~F., {Satyapal}, S., \& {Ellison}, S.~L. 2018,
  \mnras, 478, 3056, \dodoi{10.1093/mnras/sty1274}

\bibitem[{{Brinchmann} {et~al.}(2004){Brinchmann}, {Charlot}, {Heckman},
  {Kauffmann}, {Tremonti}, \& {White}}]{brinchmann2004}
{Brinchmann}, J., {Charlot}, S., {Heckman}, T.~M., {et~al.} 2004, arXiv
  e-prints, astro, \dodoi{10.48550/arXiv.astro-ph/0406220}

\bibitem[{{Brorby} {et~al.}(2014){Brorby}, {Kaaret}, \&
  {Prestwich}}]{brorby2014}
{Brorby}, M., {Kaaret}, P., \& {Prestwich}, A. 2014, \mnras, 441, 2346,
  \dodoi{10.1093/mnras/stu736}

\bibitem[{{Burke} {et~al.}(2021){Burke}, {Liu}, {Chen}, {Shen}, \&
  {Guo}}]{burke2021}
{Burke}, C.~J., {Liu}, X., {Chen}, Y.-C., {Shen}, Y., \& {Guo}, H. 2021,
  \mnras, 504, 543, \dodoi{10.1093/mnras/stab912}

\bibitem[{{Cair{\'o}s} {et~al.}(2010){Cair{\'o}s}, {Caon}, {Zurita}, {Kehrig},
  {Roth}, \& {Weilbacher}}]{cairos2010}
{Cair{\'o}s}, L.~M., {Caon}, N., {Zurita}, C., {et~al.} 2010, \aap, 520, A90,
  \dodoi{10.1051/0004-6361/201014004}

\bibitem[{{Cann} {et~al.}(2018){Cann}, {Satyapal}, {Abel}, {Ricci}, {Secrest},
  {Blecha}, \& {Gliozzi}}]{cann2018}
{Cann}, J.~M., {Satyapal}, S., {Abel}, N.~P., {et~al.} 2018, \apj, 861, 142,
  \dodoi{10.3847/1538-4357/aac64a}

\bibitem[{{Cann} {et~al.}(2020){Cann}, {Satyapal}, {Bohn}, {Sexton}, {Pfeifle},
  {Manzano-King}, {Canalizo}, {Rothberg}, {Gliozzi}, {Secrest}, \&
  {Blecha}}]{cann2020}
{Cann}, J.~M., {Satyapal}, S., {Bohn}, T., {et~al.} 2020, \apj, 895, 147,
  \dodoi{10.3847/1538-4357/ab8b64}

\bibitem[{{Cann} {et~al.}(2021){Cann}, {Satyapal}, {Rothberg}, {Canalizo},
  {Bohn}, {LaMassa}, {Matzko}, {Blecha}, {Secrest}, {Seth}, {B{\"o}ker},
  {Sexton}, {Kamal}, \& {Schmitt}}]{cann2021}
{Cann}, J.~M., {Satyapal}, S., {Rothberg}, B., {et~al.} 2021, \apjl, 912, L2,
  \dodoi{10.3847/2041-8213/abf56d}

\bibitem[{{Castro} {et~al.}(2017){Castro}, {Dors}, {Cardaci}, \&
  {H{\"a}gele}}]{castro2017}
{Castro}, C.~S., {Dors}, O.~L., {Cardaci}, M.~V., \& {H{\"a}gele}, G.~F. 2017,
  \mnras, 467, 1507, \dodoi{10.1093/mnras/stx150}

\bibitem[{{Cutri} {et~al.}(2021){Cutri}, {Wright}, {Conrow}, {Fowler},
  {Eisenhardt}, {Grillmair}, {Kirkpatrick}, {Masci}, {McCallon}, {Wheelock},
  {Fajardo-Acosta}, {Yan}, {Benford}, {Harbut}, {Jarrett}, {Lake}, {Leisawitz},
  {Ressler}, {Stanford}, {Tsai}, {Liu}, {Helou}, {Mainzer}, {Gettngs},
  {Gonzalez}, {Hoffman}, {Marsh}, {Padgett}, {Skrutskie}, {Beck}, {Papin}, \&
  {Wittman}}]{cutri2013}
{Cutri}, R.~M., {Wright}, E.~L., {Conrow}, T., {et~al.} 2021, VizieR Online
  Data Catalog, II/328

\bibitem[{{Di Matteo} {et~al.}(2005){Di Matteo}, {Springel}, \&
  {Hernquist}}]{dimatteo2005}
{Di Matteo}, T., {Springel}, V., \& {Hernquist}, L. 2005, \nat, 433, 604,
  \dodoi{10.1038/nature03335}

\bibitem[{{Dong} {et~al.}(2012){Dong}, {Greene}, \& {Ho}}]{dong2012}
{Dong}, R., {Greene}, J.~E., \& {Ho}, L.~C. 2012, \apj, 761, 73,
  \dodoi{10.1088/0004-637X/761/1/73}

\bibitem[{{Fragos} {et~al.}(2013){Fragos}, {Lehmer}, {Tremmel}, {Tzanavaris},
  {Basu-Zych}, {Belczynski}, {Hornschemeier}, {Jenkins}, {Kalogera}, {Ptak}, \&
  {Zezas}}]{fragos2013}
{Fragos}, T., {Lehmer}, B., {Tremmel}, M., {et~al.} 2013, \apj, 764, 41,
  \dodoi{10.1088/0004-637X/764/1/41}

\bibitem[{{Gabriel} {et~al.}(2004){Gabriel}, {Denby}, {Fyfe}, {Hoar}, {Ibarra},
  {Ojero}, {Osborne}, {Saxton}, {Lammers}, \& {Vacanti}}]{sas}
{Gabriel}, C., {Denby}, M., {Fyfe}, D.~J., {et~al.} 2004, in Astronomical
  Society of the Pacific Conference Series, Vol. 314, Astronomical Data
  Analysis Software and Systems (ADASS) XIII, ed. F.~{Ochsenbein}, M.~G.
  {Allen}, \& D.~{Egret}, 759

\bibitem[{{Ghosh} \& {Rana}(2021)}]{ghosh2021}
{Ghosh}, T., \& {Rana}, V. 2021, \mnras, 504, 974,
  \dodoi{10.1093/mnras/stab774}

\bibitem[{{Greene} \& {Ho}(2007)}]{greene2007}
{Greene}, J.~E., \& {Ho}, L.~C. 2007, \apj, 670, 92, \dodoi{10.1086/522082}

\bibitem[{{Greene} {et~al.}(2020){Greene}, {Strader}, \& {Ho}}]{greene2020}
{Greene}, J.~E., {Strader}, J., \& {Ho}, L.~C. 2020, \araa, 58, 257,
  \dodoi{10.1146/annurev-astro-032620-021835}

\bibitem[{{Groves} {et~al.}(2006){Groves}, {Heckman}, \&
  {Kauffmann}}]{groves2006}
{Groves}, B.~A., {Heckman}, T.~M., \& {Kauffmann}, G. 2006, \mnras, 371, 1559,
  \dodoi{10.1111/j.1365-2966.2006.10812.x}

\bibitem[{{Guseva} {et~al.}(2017){Guseva}, {Izotov}, {Fricke}, \&
  {Henkel}}]{guseva2017}
{Guseva}, N.~G., {Izotov}, Y.~I., {Fricke}, K.~J., \& {Henkel}, C. 2017, \aap,
  599, A65, \dodoi{10.1051/0004-6361/201629181}

\bibitem[{{Guseva} {et~al.}(2019){Guseva}, {Izotov}, {Fricke}, \&
  {Henkel}}]{guseva2019}
---. 2019, \aap, 624, A21, \dodoi{10.1051/0004-6361/201834935}

\bibitem[{{Hopkins} {et~al.}(2008){Hopkins}, {Hernquist}, {Cox}, \&
  {Kere{\v{s}}}}]{hopkins2008}
{Hopkins}, P.~F., {Hernquist}, L., {Cox}, T.~J., \& {Kere{\v{s}}}, D. 2008,
  \apjs, 175, 356, \dodoi{10.1086/524362}

\bibitem[{{Izotov} {et~al.}(2015){Izotov}, {Guseva}, {Fricke}, \&
  {Henkel}}]{izotov2015}
{Izotov}, Y.~I., {Guseva}, N.~G., {Fricke}, K.~J., \& {Henkel}, C. 2015,
  \mnras, 451, 2251, \dodoi{10.1093/mnras/stv1115}

\bibitem[{{Izotov} {et~al.}(2019{\natexlab{a}}){Izotov}, {Guseva}, {Fricke}, \&
  {Henkel}}]{izotov2019a}
---. 2019{\natexlab{a}}, \aap, 623, A40, \dodoi{10.1051/0004-6361/201834768}

\bibitem[{{Izotov} {et~al.}(2010){Izotov}, {Guseva}, {Fricke}, {Stasi{\'n}ska},
  {Henkel}, \& {Papaderos}}]{izotov2010}
{Izotov}, Y.~I., {Guseva}, N.~G., {Fricke}, K.~J., {et~al.} 2010, \aap, 517,
  A90, \dodoi{10.1051/0004-6361/201014390}

\bibitem[{{Izotov} {et~al.}(2006){Izotov}, {Stasi{\'n}ska}, {Meynet}, {Guseva},
  \& {Thuan}}]{izotov2006}
{Izotov}, Y.~I., {Stasi{\'n}ska}, G., {Meynet}, G., {Guseva}, N.~G., \&
  {Thuan}, T.~X. 2006, \aap, 448, 955, \dodoi{10.1051/0004-6361:20053763}

\bibitem[{{Izotov} \& {Thuan}(2008)}]{izotov2008}
{Izotov}, Y.~I., \& {Thuan}, T.~X. 2008, \apj, 687, 133, \dodoi{10.1086/591660}

\bibitem[{{Izotov} \& {Thuan}(2016)}]{izotov2016}
---. 2016, \mnras, 457, 64, \dodoi{10.1093/mnras/stv2957}

\bibitem[{{Izotov} {et~al.}(2007){Izotov}, {Thuan}, \& {Guseva}}]{izotov2007}
{Izotov}, Y.~I., {Thuan}, T.~X., \& {Guseva}, N.~G. 2007, \apj, 671, 1297,
  \dodoi{10.1086/522923}

\bibitem[{{Izotov} {et~al.}(2012{\natexlab{a}}){Izotov}, {Thuan}, \&
  {Guseva}}]{izotov2012b}
---. 2012{\natexlab{a}}, \aap, 546, A122, \dodoi{10.1051/0004-6361/201219733}

\bibitem[{{Izotov} {et~al.}(2019{\natexlab{b}}){Izotov}, {Thuan}, \&
  {Guseva}}]{izotov2019b}
---. 2019{\natexlab{b}}, \mnras, 483, 5491, \dodoi{10.1093/mnras/sty3472}

\bibitem[{{Izotov} {et~al.}(2021{\natexlab{a}}){Izotov}, {Thuan}, \&
  {Guseva}}]{izotov2021a}
---. 2021{\natexlab{a}}, \mnras, 504, 3996, \dodoi{10.1093/mnras/stab1099}

\bibitem[{{Izotov} {et~al.}(2021{\natexlab{b}}){Izotov}, {Thuan}, \&
  {Guseva}}]{izotov2021b}
---. 2021{\natexlab{b}}, \mnras, 508, 2556, \dodoi{10.1093/mnras/stab2798}

\bibitem[{{Izotov} {et~al.}(2018){Izotov}, {Thuan}, {Guseva}, \&
  {Liss}}]{izotov2018}
{Izotov}, Y.~I., {Thuan}, T.~X., {Guseva}, N.~G., \& {Liss}, S.~E. 2018,
  \mnras, 473, 1956, \dodoi{10.1093/mnras/stx2478}

\bibitem[{{Izotov} {et~al.}(2012{\natexlab{b}}){Izotov}, {Thuan}, \&
  {Privon}}]{izotov2012}
{Izotov}, Y.~I., {Thuan}, T.~X., \& {Privon}, G. 2012{\natexlab{b}}, \mnras,
  427, 1229, \dodoi{10.1111/j.1365-2966.2012.22051.x}

\bibitem[{{Jarrett} {et~al.}(2011){Jarrett}, {Cohen}, {Masci}, {Wright},
  {Stern}, {Benford}, {Blain}, {Carey}, {Cutri}, {Eisenhardt}, {Lonsdale},
  {Mainzer}, {Marsh}, {Padgett}, {Petty}, {Ressler}, {Skrutskie}, {Stanford},
  {Surace}, {Tsai}, {Wheelock}, \& {Yan}}]{jarrett2011}
{Jarrett}, T.~H., {Cohen}, M., {Masci}, F., {et~al.} 2011, \apj, 735, 112,
  \dodoi{10.1088/0004-637X/735/2/112}

\bibitem[{{Ju} {et~al.}(2022){Ju}, {Yin}, {Liu}, {Hao}, {Shao}, {Feng},
  {Riffel}, {Liu}, {Stark}, {Shen}, {Telles}, {Fern{\'a}ndez-Trincado}, {Wang},
  {Xu}, {Bizyaev}, \& {Rong}}]{ju2022}
{Ju}, M., {Yin}, J., {Liu}, R., {et~al.} 2022, arXiv e-prints,
  arXiv:2209.03298.
\newblock \doarXiv{2209.03298}

\bibitem[{{Kaaret} {et~al.}(2011){Kaaret}, {Schmitt}, \& {Gorski}}]{kaaret2011}
{Kaaret}, P., {Schmitt}, J., \& {Gorski}, M. 2011, \apj, 741, 10,
  \dodoi{10.1088/0004-637X/741/1/10}

\bibitem[{{Kauffmann} {et~al.}(2003){Kauffmann}, {Heckman}, {Tremonti},
  {Brinchmann}, {Charlot}, {White}, {Ridgway}, {Brinkmann}, {Fukugita}, {Hall},
  {Ivezi{\'c}}, {Richards}, \& {Schneider}}]{kauffmann2003}
{Kauffmann}, G., {Heckman}, T.~M., {Tremonti}, C., {et~al.} 2003, \mnras, 346,
  1055, \dodoi{10.1111/j.1365-2966.2003.07154.x}

\bibitem[{{Kelly}(2007)}]{kelly2007}
{Kelly}, B.~C. 2007, \apj, 665, 1489, \dodoi{10.1086/519947}

\bibitem[{{Kewley} {et~al.}(2001){Kewley}, {Dopita}, {Sutherland}, {Heisler},
  \& {Trevena}}]{kewley2001}
{Kewley}, L.~J., {Dopita}, M.~A., {Sutherland}, R.~S., {Heisler}, C.~A., \&
  {Trevena}, J. 2001, \apj, 556, 121, \dodoi{10.1086/321545}

\bibitem[{{Kraft} {et~al.}(2022){Kraft}, {Markevitch}, {Kilbourne}, {Adams},
  {Akamatsu}, {Ayromlou}, {Bandler}, {Bennett}, {Bhardwaj}, {Biffi},
  {Bodewits}, {Bogdan}, {Bonamente}, {Borgani}, {Branduardi-Raymont},
  {Bregman}, {Burchett}, {Cann}, {Carter}, {Chakraborty}, {Churazov}, {Crain},
  {Cumbee}, {Dave}, {DiPirro}, {Dolag}, {Bertrand Doriese}, {Drake}, {Dunn},
  {Eckart}, {Eckert}, {Ettori}, {Forman}, {Galeazzi}, {Gall}, {Gatuzz}, {Hell},
  {Hodges-Kluck}, {Jackman}, {Jahromi}, {Jennings}, {Jones}, {Kaaret},
  {Kavanagh}, {Kelley}, {Khabibullin}, {Kim}, {Koutroumpa}, {Kovacs}, {Kuntz},
  {Lin}, {Lau}, {Lee}, {Leutenegger}, {Lisse}, {Lovisari}, {McCammon},
  {McEntee}, {Mernier}, {Miller}, {Nagai}, {Negro}, {Nelson}, {Ness}, {Nulsen},
  {Ogorzalek}, {Oppenheimer}, {Oskinova}, {Patnaude}, {Pfeifle}, {Pillepich},
  {Plucinsky}, {Pooley}, {Porter}, {Randall}, {Rasia}, {Raymond}, {Ruszkowski},
  {Sakai}, {Sarkar}, {Sasaki}, {Sato}, {Schellenberger}, {Schaye},
  {Simionescu}, {Smith}, {Steiner}, {Stern}, {Su}, {Sun}, {Tremblay}, {Truong},
  {Tutt}, {Veilleux}, {Vikhlinin}, {Vladutescu-Zopp}, {Vogelsberger}, {Walker},
  {Weaver}, {Weigt}, {Werk}, {Werner}, {Wolk}, {Zhang}, {Zhang}, {Zhuravleva},
  \& {ZuHone}}]{kraft2022}
{Kraft}, R., {Markevitch}, M., {Kilbourne}, C., {et~al.} 2022, arXiv e-prints,
  arXiv:2211.09827, \dodoi{10.48550/arXiv.2211.09827}

\bibitem[{{Lamastra} {et~al.}(2009){Lamastra}, {Bianchi}, {Matt}, {Perola},
  {Barcons}, \& {Carrera}}]{lamastra2009}
{Lamastra}, A., {Bianchi}, S., {Matt}, G., {et~al.} 2009, \aap, 504, 73,
  \dodoi{10.1051/0004-6361/200912023}

\bibitem[{{Latimer} {et~al.}(2021){Latimer}, {Reines}, {Bogdan}, \&
  {Kraft}}]{latimer2021}
{Latimer}, L.~J., {Reines}, A.~E., {Bogdan}, A., \& {Kraft}, R. 2021, \apjl,
  922, L40, \dodoi{10.3847/2041-8213/ac3af6}

\bibitem[{{Lehmer} {et~al.}(2022){Lehmer}, {Eufrasio}, {Basu-Zych}, {Garofali},
  {Gilbertson}, {Mesinger}, \& {Yukita}}]{lehmer2022}
{Lehmer}, B.~D., {Eufrasio}, R.~T., {Basu-Zych}, A., {et~al.} 2022, \apj, 930,
  135, \dodoi{10.3847/1538-4357/ac63a7}

\bibitem[{{Lehmer} {et~al.}(2019){Lehmer}, {Eufrasio}, {Tzanavaris},
  {Basu-Zych}, {Fragos}, {Prestwich}, {Yukita}, {Zezas}, {Hornschemeier}, \&
  {Ptak}}]{lehmer2019}
{Lehmer}, B.~D., {Eufrasio}, R.~T., {Tzanavaris}, P., {et~al.} 2019, \apjs,
  243, 3, \dodoi{10.3847/1538-4365/ab22a8}

\bibitem[{{Lehmer} {et~al.}(2021){Lehmer}, {Eufrasio}, {Basu-Zych}, {Doore},
  {Fragos}, {Garofali}, {Kovlakas}, {Williams}, {Zezas}, \&
  {Santana-Silva}}]{lehmer2021}
{Lehmer}, B.~D., {Eufrasio}, R.~T., {Basu-Zych}, A., {et~al.} 2021, \apj, 907,
  17, \dodoi{10.3847/1538-4357/abcec1}

\bibitem[{{Mapelli} {et~al.}(2010){Mapelli}, {Ripamonti}, {Zampieri}, {Colpi},
  \& {Bressan}}]{mapelli2010}
{Mapelli}, M., {Ripamonti}, E., {Zampieri}, L., {Colpi}, M., \& {Bressan}, A.
  2010, \mnras, 408, 234, \dodoi{10.1111/j.1365-2966.2010.17048.x}

\bibitem[{{Mezcua}(2019)}]{mezcua2019}
{Mezcua}, M. 2019, Nature Astronomy, 3, 6, \dodoi{10.1038/s41550-018-0662-2}

\bibitem[{{Molina} {et~al.}(2021){Molina}, {Reines}, {Latimer}, {Baldassare},
  \& {Salehirad}}]{molina2021}
{Molina}, M., {Reines}, A.~E., {Latimer}, L.~J., {Baldassare}, V., \&
  {Salehirad}, S. 2021, \apj, 922, 155, \dodoi{10.3847/1538-4357/ac1ffa}

\bibitem[{{Mukai}(1993)}]{mukai1993}
{Mukai}, K. 1993, Legacy, 3, 21

\bibitem[{{Nandra} {et~al.}(2013){Nandra}, {Barret}, {Barcons}, {Fabian}, {den
  Herder}, {Piro}, {Watson}, {Adami}, {Aird}, {Afonso}, {Alexander},
  {Argiroffi}, {Amati}, {Arnaud}, {Atteia}, {Audard}, {Badenes}, {Ballet},
  {Ballo}, {Bamba}, {Bhardwaj}, {Stefano Battistelli}, {Becker}, {De Becker},
  {Behar}, {Bianchi}, {Biffi}, {B{\^\i}rzan}, {Bocchino}, {Bogdanov}, {Boirin},
  {Boller}, {Borgani}, {Borm}, {Bouch{\'e}}, {Bourdin}, {Bower}, {Braito},
  {Branchini}, {Branduardi-Raymont}, {Bregman}, {Brenneman}, {Brightman},
  {Br{\"u}ggen}, {Buchner}, {Bulbul}, {Brusa}, {Bursa}, {Caccianiga},
  {Cackett}, {Campana}, {Cappelluti}, {Cappi}, {Carrera}, {Ceballos},
  {Christensen}, {Chu}, {Churazov}, {Clerc}, {Corbel}, {Corral}, {Comastri},
  {Costantini}, {Croston}, {Dadina}, {D'Ai}, {Decourchelle}, {Della Ceca},
  {Dennerl}, {Dolag}, {Done}, {Dovciak}, {Drake}, {Eckert}, {Edge}, {Ettori},
  {Ezoe}, {Feigelson}, {Fender}, {Feruglio}, {Finoguenov}, {Fiore}, {Galeazzi},
  {Gallagher}, {Gandhi}, {Gaspari}, {Gastaldello}, {Georgakakis},
  {Georgantopoulos}, {Gilfanov}, {Gitti}, {Gladstone}, {Goosmann}, {Gosset},
  {Grosso}, {Guedel}, {Guerrero}, {Haberl}, {Hardcastle}, {Heinz}, {Alonso
  Herrero}, {Herv{\'e}}, {Holmstrom}, {Iwasawa}, {Jonker}, {Kaastra}, {Kara},
  {Karas}, {Kastner}, {King}, {Kosenko}, {Koutroumpa}, {Kraft}, {Kreykenbohm},
  {Lallement}, {Lanzuisi}, {Lee}, {Lemoine-Goumard}, {Lobban}, {Lodato},
  {Lovisari}, {Lotti}, {McCharthy}, {McNamara}, {Maggio}, {Maiolino}, {De
  Marco}, {de Martino}, {Mateos}, {Matt}, {Maughan}, {Mazzotta}, {Mendez},
  {Merloni}, {Micela}, {Miceli}, {Mignani}, {Miller}, {Miniutti}, {Molendi},
  {Montez}, {Moretti}, {Motch}, {Naz{\'e}}, {Nevalainen}, {Nicastro}, {Nulsen},
  {Ohashi}, {O'Brien}, {Osborne}, {Oskinova}, {Pacaud}, {Paerels}, {Page},
  {Papadakis}, {Pareschi}, {Petre}, {Petrucci}, {Piconcelli}, {Pillitteri},
  {Pinto}, {de Plaa}, {Pointecouteau}, {Ponman}, {Ponti}, {Porquet}, {Pounds},
  {Pratt}, {Predehl}, {Proga}, {Psaltis}, {Rafferty}, {Ramos-Ceja}, {Ranalli},
  {Rasia}, {Rau}, {Rauw}, {Rea}, {Read}, {Reeves}, {Reiprich}, {Renaud},
  {Reynolds}, {Risaliti}, {Rodriguez}, {Rodriguez Hidalgo}, {Roncarelli},
  {Rosario}, {Rossetti}, {Rozanska}, {Rovilos}, {Salvaterra}, {Salvato}, {Di
  Salvo}, {Sanders}, {Sanz-Forcada}, {Schawinski}, {Schaye}, {Schwope},
  {Sciortino}, {Severgnini}, {Shankar}, {Sijacki}, {Sim}, {Schmid}, {Smith},
  {Steiner}, {Stelzer}, {Stewart}, {Strohmayer}, {Str{\"u}der}, {Sun}, {Takei},
  {Tatischeff}, {Tiengo}, {Tombesi}, {Trinchieri}, {Tsuru}, {Ud-Doula},
  {Ursino}, {Valencic}, {Vanzella}, {Vaughan}, {Vignali}, {Vink}, {Vito},
  {Volonteri}, {Wang}, {Webb}, {Willingale}, {Wilms}, {Wise}, {Worrall},
  {Young}, {Zampieri}, {In't Zand}, {Zane}, {Zezas}, {Zhang}, \&
  {Zhuravleva}}]{nandra2013}
{Nandra}, K., {Barret}, D., {Barcons}, X., {et~al.} 2013, arXiv e-prints,
  arXiv:1306.2307.
\newblock \doarXiv{1306.2307}

\bibitem[{{Panessa} {et~al.}(2006){Panessa}, {Bassani}, {Cappi}, {Dadina},
  {Barcons}, {Carrera}, {Ho}, \& {Iwasawa}}]{panessa2006}
{Panessa}, F., {Bassani}, L., {Cappi}, M., {et~al.} 2006, \aap, 455, 173,
  \dodoi{10.1051/0004-6361:20064894}

\bibitem[{{Pettini} \& {Pagel}(2004)}]{pettini2004}
{Pettini}, M., \& {Pagel}, B. E.~J. 2004, \mnras, 348, L59,
  \dodoi{10.1111/j.1365-2966.2004.07591.x}

\bibitem[{{Pilyugin} \& {Thuan}(2005)}]{pilyugin2005}
{Pilyugin}, L.~S., \& {Thuan}, T.~X. 2005, \apj, 631, 231,
  \dodoi{10.1086/432408}

\bibitem[{{Prestwich} {et~al.}(2013){Prestwich}, {Tsantaki}, {Zezas},
  {Jackson}, {Roberts}, {Foltz}, {Linden}, \& {Kalogera}}]{prestwich2013}
{Prestwich}, A.~H., {Tsantaki}, M., {Zezas}, A., {et~al.} 2013, \apj, 769, 92,
  \dodoi{10.1088/0004-637X/769/2/92}

\bibitem[{{Reefe} {et~al.}(2022){Reefe}, {Satyapal}, {Sexton}, {Doan},
  {Secrest}, \& {Cann}}]{reefe2022a}
{Reefe}, M., {Satyapal}, S., {Sexton}, R.~O., {et~al.} 2022, \apj, 936, 140,
  \dodoi{10.3847/1538-4357/ac8981}

\bibitem[{{Reines} \& {Comastri}(2016)}]{reines2016}
{Reines}, A.~E., \& {Comastri}, A. 2016, \pasa, 33, e054,
  \dodoi{10.1017/pasa.2016.46}

\bibitem[{{Salim} {et~al.}(2007){Salim}, {Rich}, {Charlot}, {Brinchmann},
  {Johnson}, {Schiminovich}, {Seibert}, {Mallery}, {Heckman}, {Forster},
  {Friedman}, {Martin}, {Morrissey}, {Neff}, {Small}, {Wyder}, {Bianchi},
  {Donas}, {Lee}, {Madore}, {Milliard}, {Szalay}, {Welsh}, \& {Yi}}]{salim2007}
{Salim}, S., {Rich}, R.~M., {Charlot}, S., {et~al.} 2007, \apjs, 173, 267,
  \dodoi{10.1086/519218}

\bibitem[{{Salvato} {et~al.}(2022){Salvato}, {Wolf}, {Dwelly}, {Georgakakis},
  {Brusa}, {Merloni}, {Liu}, {Toba}, {Nandra}, {Lamer}, {Buchner}, {Schneider},
  {Freund}, {Rau}, {Schwope}, {Nishizawa}, {Klein}, {Arcodia}, {Comparat},
  {Musiimenta}, {Nagao}, {Brunner}, {Malyali}, {Finoguenov}, {Anderson},
  {Shen}, {Ibarra-Medel}, {Trump}, {Brandt}, {Urry}, {Rivera}, {Krumpe},
  {Urrutia}, {Miyaji}, {Ichikawa}, {Schneider}, {Fresco}, {Boller}, {Haase},
  {Brownstein}, {Lane}, {Bizyaev}, \& {Nitschelm}}]{salvato2022}
{Salvato}, M., {Wolf}, J., {Dwelly}, T., {et~al.} 2022, \aap, 661, A3,
  \dodoi{10.1051/0004-6361/202141631}

\bibitem[{{Secrest} {et~al.}(2020){Secrest}, {Ellison}, {Satyapal}, \&
  {Blecha}}]{secrest2020}
{Secrest}, N.~J., {Ellison}, S.~L., {Satyapal}, S., \& {Blecha}, L. 2020,
  \mnras, 499, 2380, \dodoi{10.1093/mnras/staa1692}

\bibitem[{{Secrest} {et~al.}(2015){Secrest}, {Satyapal}, {Gliozzi}, {Rothberg},
  {Ellison}, {Mowry}, {Rosenberg}, {Fischer}, \& {Schmitt}}]{secrest2015}
{Secrest}, N.~J., {Satyapal}, S., {Gliozzi}, M., {et~al.} 2015, \apj, 798, 38,
  \dodoi{10.1088/0004-637X/798/1/38}

\bibitem[{{Simmonds} {et~al.}(2016){Simmonds}, {Bauer}, {Thuan}, {Izotov},
  {Stern}, \& {Harrison}}]{simmonds2016}
{Simmonds}, C., {Bauer}, F.~E., {Thuan}, T.~X., {et~al.} 2016, \aap, 596, A64,
  \dodoi{10.1051/0004-6361/201629310}

\bibitem[{{Steffen} {et~al.}(2006){Steffen}, {Strateva}, {Brandt}, {Alexander},
  {Koekemoer}, {Lehmer}, {Schneider}, \& {Vignali}}]{steffen2006}
{Steffen}, A.~T., {Strateva}, I., {Brandt}, W.~N., {et~al.} 2006, \aj, 131,
  2826, \dodoi{10.1086/503627}

\bibitem[{{Stern} {et~al.}(2012){Stern}, {Assef}, {Benford}, {Blain}, {Cutri},
  {Dey}, {Eisenhardt}, {Griffith}, {Jarrett}, {Lake}, {Masci}, {Petty},
  {Stanford}, {Tsai}, {Wright}, {Yan}, {Harrison}, \& {Madsen}}]{stern2012}
{Stern}, D., {Assef}, R.~J., {Benford}, D.~J., {et~al.} 2012, \apj, 753, 30,
  \dodoi{10.1088/0004-637X/753/1/30}

\bibitem[{{Storchi-Bergmann} {et~al.}(1998){Storchi-Bergmann}, {Schmitt},
  {Calzetti}, \& {Kinney}}]{storchibergmann1998}
{Storchi-Bergmann}, T., {Schmitt}, H.~R., {Calzetti}, D., \& {Kinney}, A.~L.
  1998, \aj, 115, 909, \dodoi{10.1086/300242}

\bibitem[{{Taylor}(2005)}]{topcat}
{Taylor}, M.~B. 2005, in Astronomical Society of the Pacific Conference Series,
  Vol. 347, Astronomical Data Analysis Software and Systems XIV, ed.
  P.~{Shopbell}, M.~{Britton}, \& R.~{Ebert}, 29

\bibitem[{{Thuan} {et~al.}(2016){Thuan}, {Goehring}, {Hibbard}, {Izotov}, \&
  {Hunt}}]{thuan2016}
{Thuan}, T.~X., {Goehring}, K.~M., {Hibbard}, J.~E., {Izotov}, Y.~I., \&
  {Hunt}, L.~K. 2016, \mnras, 463, 4268, \dodoi{10.1093/mnras/stw2259}

\bibitem[{{Thuan} {et~al.}(2022){Thuan}, {Guseva}, \& {Izotov}}]{thuan2022}
{Thuan}, T.~X., {Guseva}, N.~G., \& {Izotov}, Y.~I. 2022, \mnras, 516, L81,
  \dodoi{10.1093/mnrasl/slac095}

\bibitem[{{Thygesen} {et~al.}(2023){Thygesen}, {Plotkin}, {Soria}, {Reines},
  {Greene}, {Anderson}, {Baldassare}, {Owens}, {Urquhart}, {Gallo},
  {Miller-Jones}, {Paul}, \& {Rollings}}]{thygesen2023}
{Thygesen}, E., {Plotkin}, R.~M., {Soria}, R., {et~al.} 2023, \mnras, 519,
  5848, \dodoi{10.1093/mnras/stad002}

\bibitem[{{Tremonti} {et~al.}(2004){Tremonti}, {Heckman}, {Kauffmann},
  {Brinchmann}, {Charlot}, {White}, {Seibert}, {Peng}, {Schlegel}, {Uomoto},
  {Fukugita}, \& {Brinkmann}}]{tremonti2004}
{Tremonti}, C.~A., {Heckman}, T.~M., {Kauffmann}, G., {et~al.} 2004, \apj, 613,
  898, \dodoi{10.1086/423264}

\bibitem[{{Volonteri}(2010)}]{volonteri2010}
{Volonteri}, M. 2010, \aapr, 18, 279, \dodoi{10.1007/s00159-010-0029-x}

\bibitem[{{Willingale} {et~al.}(2013){Willingale}, {Starling}, {Beardmore},
  {Tanvir}, \& {O'Brien}}]{willingale2013}
{Willingale}, R., {Starling}, R.~L.~C., {Beardmore}, A.~P., {Tanvir}, N.~R., \&
  {O'Brien}, P.~T. 2013, \mnras, 431, 394, \dodoi{10.1093/mnras/stt175}

\bibitem[{{Zou} {et~al.}(2023){Zou}, {Brandt}, {Ni}, {Zhu}, {Alexander},
  {Bauer}, {Chen}, {Luo}, {Sun}, {Vignali}, {Vito}, {Xue}, \& {Yan}}]{zou2023}
{Zou}, F., {Brandt}, W.~N., {Ni}, Q., {et~al.} 2023, \apj, 950, 136,
  \dodoi{10.3847/1538-4357/acce39}

\end{thebibliography}
\bibliographystyle{aasjournal}



\end{document}